\documentclass[12pt,english,extra,safe=none,linenumbers]{gji}
\usepackage{lmodern}
\usepackage{helvet}
\usepackage{courier}

\usepackage[T1]{fontenc}
\usepackage[latin9]{inputenc}
\usepackage[letterpaper]{geometry}
\usepackage{color}
\usepackage{babel}
\usepackage{units}
\usepackage{amsmath}
\usepackage{amssymb}
\usepackage{graphicx}
\usepackage[unicode=true,pdfusetitle,
 bookmarks=true,bookmarksnumbered=false,bookmarksopen=false,
 breaklinks=false,pdfborder={0 0 0},pdfborderstyle={},backref=false,colorlinks=true]
 {hyperref}

\makeatletter
\title[]{Physically motivated moment-tensor decomposition for mining-induced seismicity}
\author[A. Rigby]{Alex Rigby$^1$ \\$^1$ Institute of Mine Seismology, 10 Church Street, Kingston, 7050, Australia}
\definecolor{orange}{RGB}{255, 134, 13}
\definecolor{purple}{RGB}{141, 29, 117}

\usepackage{placeins}
\@ifundefined{showcaptionsetup}{}{%
 \PassOptionsToPackage{caption=false}{subfig}}
\usepackage{subfig}
\makeatother

\begin{document}
\maketitle
\begin{abstract}
Compared to existing schemes, the decomposition of moment tensors for mining-induced seismic events into closing-crack and double-couple components has the advantage that each can be interpreted in terms of a physical source process (excavation convergence and slip/shear, respectively). Obviously, not every moment tensor permits such a decomposition, and we translate existing bounds on those that do to the Hudson source-type plot. For moment tensors falling within these bounds, it has previously been noted that there will be a infinite set of possible decompositions in general. We derive an implicit equation defining this set and suggest physically motivated criteria that can be used to select a single decomposition from it. Furthermore, for moment tensors falling outside the source-type bounds, we present a simple geometric scheme for determining the closest moment tensor within them (which can then be decomposed). To demonstrate the methods developed in this paper, we apply them to two catalogues of mining-induced seismicity.
\end{abstract}

\section{Introduction}

\label{sec:Introduction}

The moment tensor gives a compact description of the low-frequency radiation of a seismic source. Interpretation of a given moment tensor in terms of a physical source process can be aided by decomposing it into various components. The standard approach is to first decompose the moment tensor into isotropic and deviatoric components. This deviatoric component is then typically further decomposed into some combination of double-couple (DC) and/or compensated linear-vector dipole (CLVD) components. This second step is not unique, with a number of methodologies having been published based on different assumptions and aims \citep{julian1998non,vavryvcuk2015moment}. A widely adopted procedure, which was first proposed by \citet{knopoff1970compensated}, is to decompose into DC and CLVD components with aligned $P$- or $T$-axes. However, interpretation of such decompositions can be difficult: The coupling of DC and CLVD orientations means that even if there is a significant DC component, its nodal planes need not align with the actual plane of slip/shear in the rockmass. Furthermore, there is no obvious source process that would yield either isotropic or CLVD content in isolation (at least in the context of mining-induced seismicity). 

As proposed by \citet{ryder1988excess}, seismic sources at mines can be broadly split into two main categories. The first of these is slip/shear in the rockmass, which can be described by a DC moment tensor. The second is ``crushing'' failure in the rockmass near a mining void. The low-frequency radiation of these so-called crush-type sources, which is dominated by the convergence of the surrounding rockmass into the excavation, can be described by the closing-crack moment tensor \citep{malovichko2022description}. As we discuss in Section \ref{sec:Preliminaries}, this categorisation motivates the development of a moment-tensor decomposition into DC and closing-crack components without any restriction on their relative orientation. Although the physical motivations are different, this is essentially equivalent mathematically to the general crack plus double-couple (CDC) decomposition considered by \citet{tape2013classical} into DC and opening/tensile-crack components. For clarity, we will refer to decompositions into closing-crack and DC components as closing-CDC decompositions. 

While physically attractive, there are complications that must be overcome for closing-CDC decompositions to be determined routinely for mining-induce seismicity. The first of these is that a closing-CDC decomposition cannot be determined for every possible moment tensor. Bounds on the moment-tensor eigenvalues that permit a decomposition have been given by \citet{tape2013classical} in terms of the lune. In Section \ref{sec:Bounds}, we translate these bounds to the more commonly used source-type plot of \citet{hudson1989source}. For moment tensors falling outside these bounds, we provide a simple geometric procedure in Section \ref{sec:Approximate-decomposition} for determining the closest closing-CDC tensor based on results of \citet{tape2012geometric,tape2013classical}.

The second complication is that there are, in general, infinite possible closing-CDC decompositions for a moment tensor falling within these bounds \citep{tape2013classical}. Intuitively, this is a result of the decomposition being a function of seven parameters, while the moment tensor only provides us with six. Selection of a single decomposition requires the inclusion of an additional constraint. In Section \ref{sec:Decomposition-approach}, we propose several physically motivated methods for imparting this constraint based on knowledge of excavation geometry, stress, or geology at the source. Furthermore, we apply these methods in Section \ref{sec:Example-applications} to two catalogues of mining-induced events, where we also show how the resulting decompositions can be interpreted in terms of physical source processes.

\section{Preliminaries}

\label{sec:Preliminaries}

Mining-induced seismic events are typically classified as being slip/shear-type or crush-type \citep{ryder1988excess}. We will give an overview these source types in this section and define a moment-tensor decomposition based on them.

\subsection{Slip/shear-type sources}

The creation of mining voids results in the concentration and redistribution of stresses, which can result in slip along previously clamped structures/weaknesses or even in the shearing of intact rock \citep{ortlepp1997rock}. Assuming this episode of slip occurs along a plane relatively far from significant excavations, the radiation can be described by the traditional DC moment tensor
\begin{equation}
\mathbf{M}^{D}=M_{D}\mathbf{R}^{D}\left(\begin{array}{ccc}
1 & 0 & 0\\
0 & 0 & 0\\
0 & 0 & -1
\end{array}\right)(\mathbf{R}^{D})^{T},\label{eq:dc-tensor-1}
\end{equation}
where $M_{D}=\lVert\mathbf{M}^{D}\rVert/\sqrt{2}=\sqrt{\sum_{ij}M_{ij}^{D}/2}$ is the scalar moment of $\mathbf{M}^{D}$ and $\mathbf{R}^{D}$ is a rotation matrix. The orientation of $\mathbf{M}^{D}$ can be described in terms of the strike $\phi_{s}$ and dip $\delta$ of the slip plane and the rake angle $\lambda$, which gives the direction of slip in this plane \citep{Aki-Richards-2009}. Explicitly, in a north-east-up coordinate system, the components of $\text{\ensuremath{\mathbf{M}^{D}}}$ can be written as
\begin{align}
M_{\mathrm{NN}}^{D} & =M_{D}\left(-\sin\delta\sin2\phi_{s}\cos\lambda-\sin2\delta\sin\lambda\sin^{2}\phi_{s}\right),\nonumber \\
M_{\mathrm{EE}}^{D} & =M_{D}\left(\sin\delta\sin2\phi_{s}\cos\lambda-\sin2\delta\sin\lambda\cos^{2}\phi_{s}\right),\nonumber \\
M_{\mathrm{UU}}^{D} & =M_{D}\sin2\delta\sin\lambda,\nonumber \\
M_{\mathrm{NE}}^{D} & =M_{D}\left(\sin\delta\cos\lambda\cos2\phi_{s}+\sin2\delta\sin\lambda\sin\phi_{s}\cos\phi_{s}\right),\nonumber \\
M_{\mathrm{NU}}^{D} & =M_{D}\left(\sin\lambda\sin\phi_{s}\cos2\delta+\cos\delta\cos\lambda\cos\phi_{s}\right),\nonumber \\
M_{\mathrm{EU}}^{D} & =M_{D}\left(-\sin\lambda\cos2\delta\cos\phi_{s}+\sin\phi_{s}\cos\delta\cos\lambda\right).
\end{align}

\subsection{Crush-type sources}

\label{subsec:Crush-type-sources}

Stress concentrations resulting from the creation of mining voids can also lead to ``crushing'' failure of rock that is in close proximity to the excavation boundary \citep{ryder1988excess}. In such cases, failure is not concentrated along some predominant plane or structure, but is rather composed of a number of small-scale episodes of shear/tensile rupture (stress fracturing). Importantly, the low-frequency radiation (that is, at wavelengths larger than the diameter of the excavation) of a crush-type source is not dominated by this small-scale behaviour of the rockmass. Instead, it is dominated by the overall convergence of the surrounding rockmass into the excavation \citep{Malovichko-2020,malovichko2022description}. 

\begin{figure}
\begin{centering}
\subfloat[]{\begin{centering}
\includegraphics[bb=0bp 0bp 306bp 129bp,scale=0.7]{ 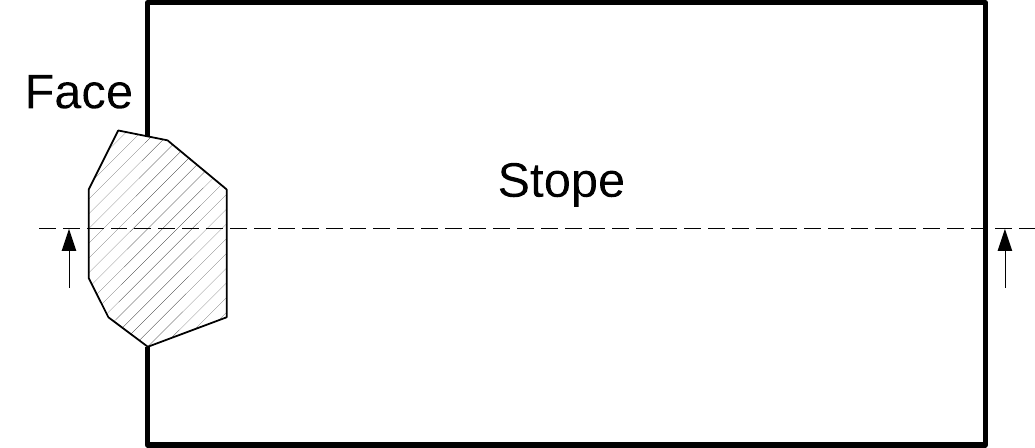}
\par\end{centering}
}
\par\end{centering}
\centering{}\subfloat[]{\begin{centering}
\includegraphics[bb=0bp 0bp 254bp 126bp,scale=0.7]{ 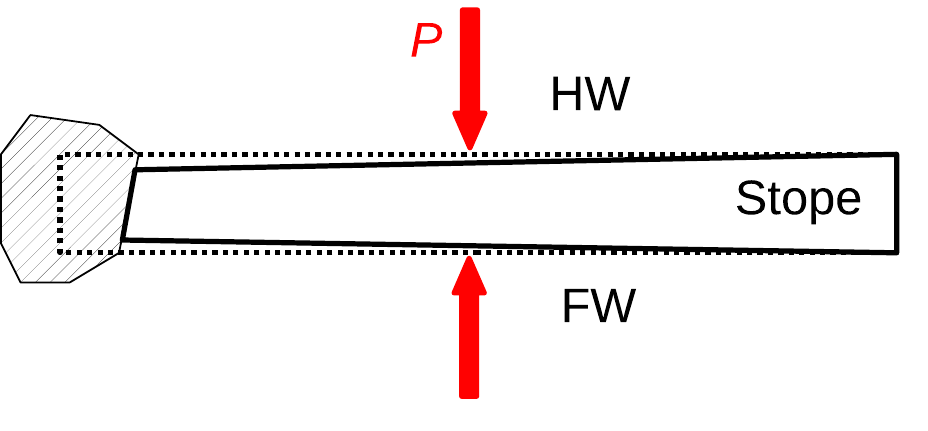}
\par\end{centering}
}\caption{(a) Plan view of a horizontal tabular stope. The hashed region indicates a volume of dynamically fractured rock at the face. (b) Section view showing the resulting convergence/closure of the footwall (FW) and hanging wall (HW) to give the post-event stope profile (solid line). The pre-event profile of the stope is given by the overlaid dotted line. Also shown is the expected $P$-axis direction.\label{fig:(a)-Plan-view}}
\end{figure}

An illustration of such a source is shown in Fig. \ref{fig:(a)-Plan-view}, where failure of a volume of rock occurs at the highly stressed face of a tabular (planar) stope. In this case, the low-frequency radiation is dominated by the convergence of the footwall and hanging wall into the stope. Given the excavation's geometry, this radiation can be described in terms of the moment tensor for a closing crack
\begin{equation}
\mathbf{M}^{K^{-}}=\alpha_{\nu}M_{K}\mathbf{R}^{K}\left(\begin{array}{ccc}
-\nu & 0 & 0\\
0 & -\nu & 0\\
0 & 0 & \nu-1
\end{array}\right)(\mathbf{R}^{K})^{T},\label{eq:crack-tensor-1}
\end{equation}
where we define $\alpha_{\nu}=2/\sqrt{4\nu^{2}+2(\nu-1)^{2}}$, $M_{K}$ is the scalar moment, $\mathbf{R}^{K}$ is an rotation matrix, and $\nu$ is the Poisson's ratio of the rockmass \citep{wong1991implosional,walter1997seismic,ford2008source}. The orientation of $\mathbf{M}^{K^{-}}$ is defined entirely by that of its $P$-axis, which will lie orthogonal to the plane of the excavation. In terms of the azimuth $\phi$ and plunge $\beta$ of the $P$-axis, we can write the components of $\mathbf{M}^{K^{-}}$ as
\begin{align}
M_{\mathrm{NN}}^{K^{-}} & =\alpha_{\nu}M_{K}\left[(2\nu-1)\cos^{2}\phi\cos^{2}\beta-\nu\right],\nonumber \\
M_{\mathrm{EE}}^{K^{-}} & =\alpha_{\nu}M_{K}\left[(2\nu-1)\sin^{2}\phi\cos^{2}\beta-\nu\right],\nonumber \\
M_{\mathrm{UU}}^{K^{-}} & =\alpha_{\nu}M_{K}\left[(2\nu-1)\sin^{2}\beta-\nu\right],\nonumber \\
M_{\mathrm{NE}}^{K^{-}} & =\alpha_{\nu}M_{K}(2\nu-1)\sin\phi\cos\phi\cos^{2}\beta,\nonumber \\
M_{\mathrm{NU}}^{K^{-}} & =\alpha_{\nu}M_{K}(1-2\nu)\cos\phi\sin\beta\cos\beta,\nonumber \\
M_{\mathrm{EU}}^{K^{-}} & =\alpha_{\nu}M_{K}(1-2\nu)\sin\phi\sin\beta\cos\beta.\label{eq:closing-crack-components}
\end{align}

\begin{figure}
\centering{}\includegraphics[bb=0bp 0bp 253bp 197bp,scale=0.7]{ 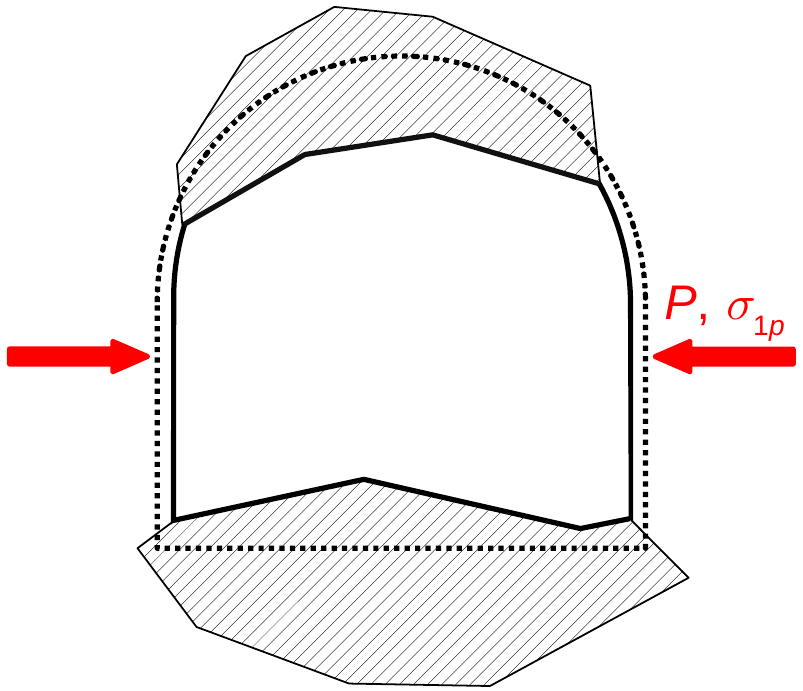}\caption{Section view of a horizontally loaded tunnel that has undergone dynamic stress fracturing in the back and floor as indicated by the hashed regions. The pre- and post-event profiles of the tunnel are given by the dotted and solid lines, respectively. Also shown is the expected $P$-axis direction, which corresponds to the direction of $\sigma_{1p}$ (the maximum in-plane loading).\label{fig:Section-view-of}}
\end{figure}

Usefully, the closing-crack model also gives a good approximation of the radiation from crush-type sources associated with different excavation geometries. This includes dynamic stress fracturing around tunnels as considered by \citet{malovichko2022description}. An illustration of such failure in the back (roof) and floor of a horizontally loaded tunnel is shown in Fig. \ref{fig:Section-view-of}. More generally, this failure will lie primarily in regions of high stress concentration. These regions occur near the excavation's surface around the direction orthogonal to maximum in-plane loading, which we call $\sigma_{1p}$. Radiation is largely controlled by resulting convergence of the surrounding rockmass into the excavation, and the $P$-axis will be approximately co-oriented with the direction of $\sigma_{1p}$.

\subsection{Source model}

\label{sec:Source-model}

\begin{figure}
\centering{}\includegraphics[bb=90bp 0bp 403bp 204bp,clip,scale=0.7]{ 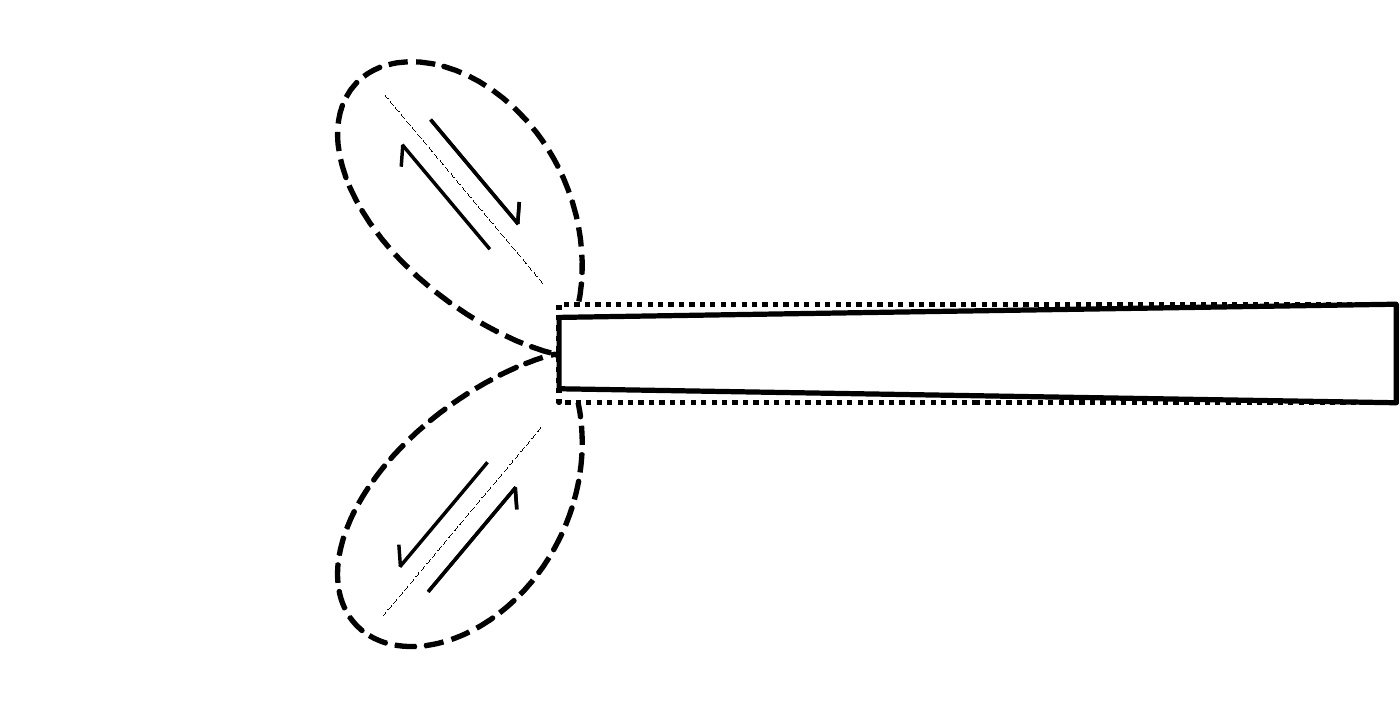}\caption{Section view of shearing ahead of the face of a tabular stope, where there are lobes of high shear stress (dashed lines). The pre- and post-event stope profiles are given by the dotted and solid lines, respectively. \label{fig:Section-view-of-1}}
\end{figure}

The slip/shear- and crush-type sources outlined above can be considered as two extremes of a spectrum of possible source processes. In practice, there are sources that combine aspects of both. For example, stress fracturing around an excavation will cause a redistribution of stress that may result in near-simultaneous slip along some nearby structure. Conversely, the dynamic stress waves from slip along a structure may result in the triggering of stress fracturing around a nearby excavation. More generally, the response (convergence) of an excavation near an episode of slip/shear will contribute to the resulting radiation even if there isn't directly induced/triggered failure around it \citep{Sileny-2001}. An example of where this can be expected is shown in Fig. \ref{fig:Section-view-of-1}, where lobes of high shear stress induce shear failure ahead of the face of a tabular stope in either the footwall or hanging wall.

We can expect the moment tensor for a source composed of an episode of slip/shear and/or excavation convergence to be a sum
\begin{equation}
\mathbf{M}=\mathbf{M}^{K^{-}}+\mathbf{M}^{D}\label{eq:source-model}
\end{equation}
of closing-crack and DC moment tensors as defined in eqs. \eqref{eq:crack-tensor-1} and \eqref{eq:dc-tensor-1}, respectively. Mathematically, this is almost identical to the general crack plus double-couple (CDC) model considered by \citep{tape2013classical}, with the only difference being that they considered a tensile/opening crack. We emphasise this difference by referring to eq. \eqref{eq:source-model} as the closing-CDC model. Furthermore, we make use of ``$-$'' superscripts for negative/closing properties (such as in $\mathbf{M}^{K^{-}}$) where necessary to differentiate from the positive/opening ones considered in Appendix \ref{sec:Arbitrary-crack}.

\section{Source-type bounds}

\label{sec:Bounds}

We can hope to gain insight into the nature of a mining-induced source by decomposing an inferred moment tensor into the closing-crack and DC components of eq. \eqref{eq:source-model}. Obviously, not every possible moment tensor can be decomposed in this manner. Bounds on those that can have previously been derived by \citet{tape2013classical}, which we present here with slight modification to account for our use of a closing rather than opening crack. We also translate these bounds to the source-type plot of \citet{hudson1989source}.

\subsection{Lune source-type plot}

\label{subsec:Lune-source-type-plot}

\begin{figure}
\begin{centering}
\subfloat[\label{fig:lune-region}]{\begin{centering}
\includegraphics[bb=0bp 0bp 224bp 235bp,scale=0.75]{ 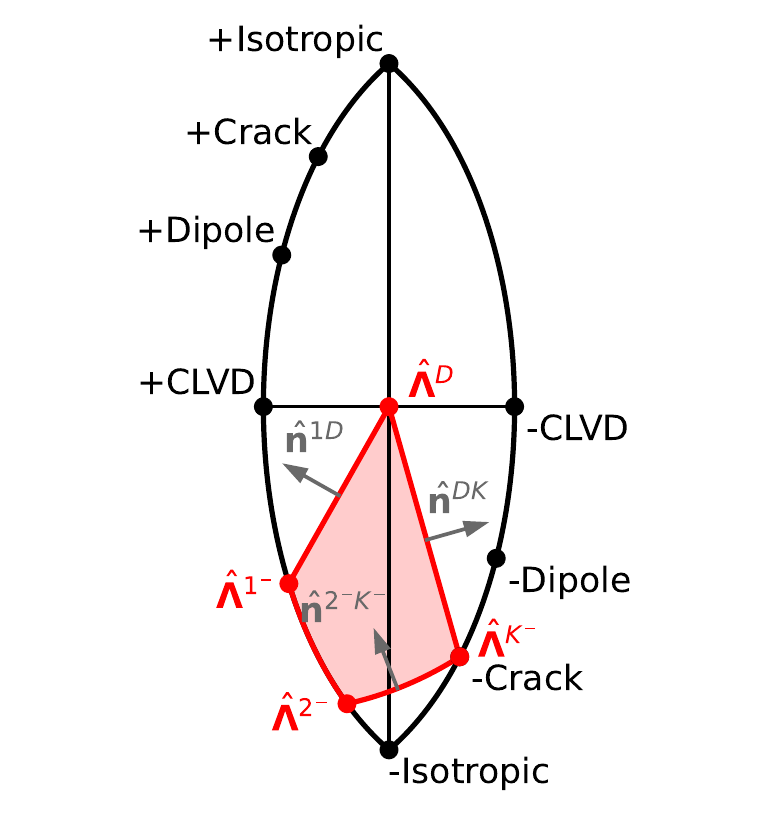}
\par\end{centering}
}
\par\end{centering}
\centering{}\subfloat[\label{fig:hudson-region}]{\begin{centering}
\hspace{-25bp}\includegraphics[bb=35bp 0bp 224bp 235bp,scale=0.75]{ 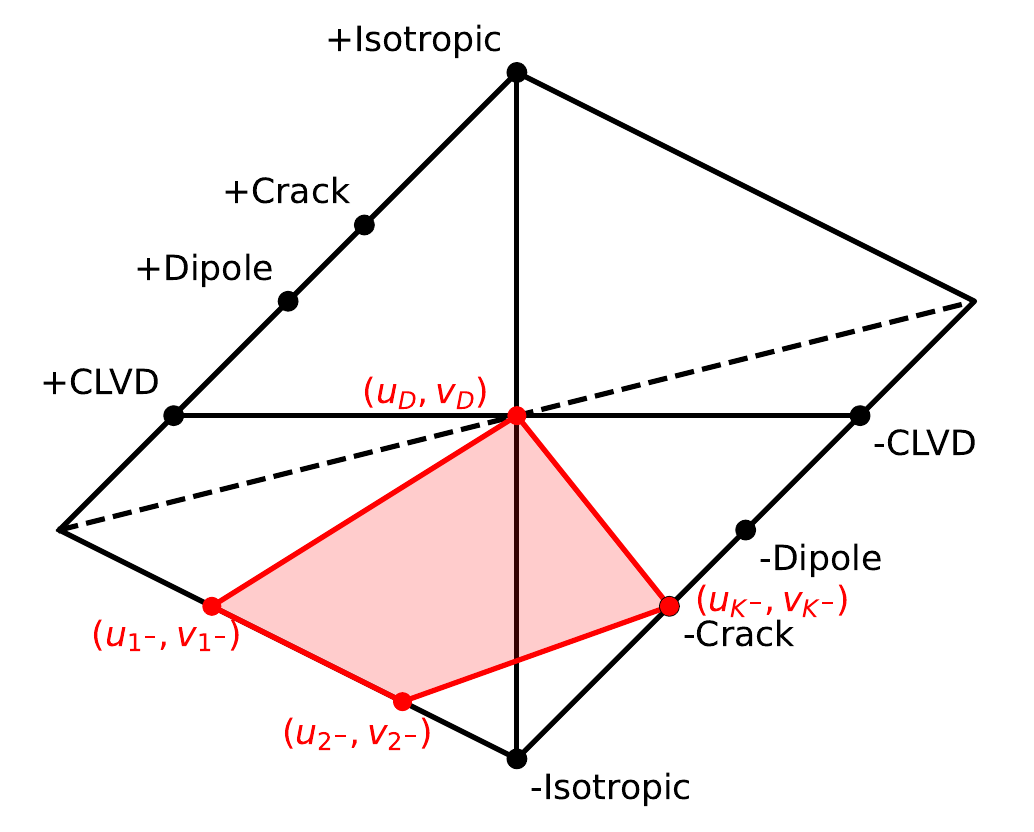}
\par\end{centering}
}\caption{(a) Region on the fundamental lune for which corresponding moment tensors permit a closing-CDC decomposition with $\nu=0.25$. (b) Corresponding region on the Hudson plot.\label{fig:Region-on-the}}
\end{figure}

We write the eigenvalues corresponding to the $T$-, $B$-, and $P$-axes of a given moment tensor $\mathbf{M}$ as the triple $\boldsymbol{\Lambda}=(\lambda_{1},\lambda_{2},\lambda_{3})$ . The set of all possible triples is the wedge
\begin{equation}
\mathbb{W}=\{\boldsymbol{\Lambda}\in\mathbb{R}^{3}:\lambda_{1}\geq\lambda_{2}\geq\lambda_{3}\},
\end{equation}
and its projection onto the unit sphere $\mathbb{S}$ is the fundamental lune $\mathbb{L}=p_{\mathbb{S}}(\mathbb{W})$, where 

\begin{equation}
p_{\mathbb{S}}(\boldsymbol{\Lambda})=\frac{\boldsymbol{\Lambda}}{\lVert\boldsymbol{\Lambda}\rVert}=\frac{\boldsymbol{\Lambda}}{\sqrt{\lambda_{1}^{2}+\lambda_{2}^{2}+\lambda_{3}^{2}}}=\hat{\boldsymbol{\Lambda}}.
\end{equation}
$\mathbf{M}$ will have a closing-CDC decomposition if its lune point $\hat{\boldsymbol{\Lambda}}(\mathbf{M})$ falls within the region $\mathbb{L}_{\mathrm{CDC^{-}}}\subset\mathbb{L}$ shown in Fig. \ref{fig:lune-region} \citep{tape2013classical}. The vertices of this region are the points
\begin{align}
\hat{\boldsymbol{\Lambda}}^{D} & =\hat{\boldsymbol{\Lambda}}(\mathbf{M}^{D})=p_{\mathbb{S}}(1,0,-1),\nonumber \\
\hat{\boldsymbol{\Lambda}}^{K^{-}} & =\hat{\boldsymbol{\Lambda}}(\mathbf{M}^{K^{-}})=p_{\mathbb{S}}(-\nu,-\nu,\nu-1),\nonumber \\
\hat{\boldsymbol{\Lambda}}^{1^{-}} & =p_{\mathbb{S}}(1-3\nu,\nu-1,\nu-1),\nonumber \\
\hat{\boldsymbol{\Lambda}}^{2^{-}} & =p_{\mathbb{S}}(-2\nu,-1,-1).\label{eq:lune-points}
\end{align}
The edges (boundary) of $\mathbb{L}_{\mathrm{CDC^{-}}}$ are segments of great circles joining these vertices, which can be defined by planes through the origin with the unit normal vectors
\begin{align}
\hat{\mathbf{n}}^{1D} & =p_{\mathbb{S}}(\hat{\boldsymbol{\Lambda}}^{1^{-}}\times\hat{\boldsymbol{\Lambda}}^{D})=p_{\mathbb{S}}(1-\nu,-2\nu,1-\nu),\nonumber \\
\hat{\mathbf{n}}^{DK} & =p_{\mathbb{S}}(\hat{\boldsymbol{\Lambda}}^{D}\times\hat{\boldsymbol{\Lambda}}^{K^{-}})=p_{\mathbb{S}}(-\nu,1,-\nu),\nonumber \\
\hat{\mathbf{n}}^{2^{-}K^{-}} & =p_{\mathbb{S}}(\hat{\boldsymbol{\Lambda}}^{2^{-}}\times\hat{\boldsymbol{\Lambda}}^{K^{-}})=p_{\mathbb{S}}(1,-\nu,-\nu).\label{eq:cdc-normals}
\end{align}
In terms of these normals, the closing-CDC region is simply
\begin{align}
\mathbb{L}_{\mathrm{CDC^{-}}} & =\{\hat{\boldsymbol{\Lambda}}\in\mathbb{L}:\hat{\mathbf{n}}^{1D}\cdot\hat{\boldsymbol{\Lambda}}\leq0,\nonumber \\
 & \phantom{=\{}\hat{\mathbf{n}}^{DK}\cdot\hat{\boldsymbol{\Lambda}}\leq0,\hat{\mathbf{n}}^{2^{-}K^{-}}\cdot\hat{\boldsymbol{\Lambda}}\geq0\}.\label{eq:cdc-region-def}
\end{align}

Some insight into the closing-CDC region can be obtained by considering the sources that lie on its boundary. For a source as defined in eq. \eqref{eq:source-model}, the boundary segments given in eq. \eqref{eq:cdc-normals} correspond to cases where the closing-crack $P$-axis is aligned with the different principal axes of the DC component: 
\begin{itemize}
\item If aligned with the DC $P$-axis, then $\hat{\mathbf{n}}^{DK}\cdot\hat{\boldsymbol{\Lambda}}=0$; that is, $\hat{\boldsymbol{\Lambda}}$ lies on the arc joining $\hat{\boldsymbol{\Lambda}}^{D}$ and $\hat{\boldsymbol{\Lambda}}^{K^{-}}$. 
\item If aligned with the DC $T$-axis, then there are two possibilities depending on the ratio $M_{D}/M_{K}$ of the DC and closing-crack scalar moments. In particular, $\hat{\mathbf{n}}^{DK}\cdot\hat{\boldsymbol{\Lambda}}=0$ if $M_{D}/M_{K}\geq(1-2\nu)\alpha_{v}$, while $\hat{\mathbf{n}}^{2^{-}K^{-}}\cdot\hat{\boldsymbol{\Lambda}}=0$ if $M_{D}/M_{K}\leq(1-2\nu)\alpha_{v}$.
\item If aligned with the DC $B$-axis, then $\hat{\mathbf{n}}^{1D}\cdot\hat{\boldsymbol{\Lambda}}=0$ for $M_{D}/M_{K}\geq(1-2\nu)\alpha_{v}$. $\hat{\boldsymbol{\Lambda}}$ does not lie on the boundary if $M_{D}/M_{K}<(1-2\nu)\alpha_{v}$.
\end{itemize}

\subsection{Hudson source-type plot}

\label{subsec:Hudson}

While there are advantages to plotting source types on the lune \citep{tape2012geometric,tape2012geometric-comparison,tape2019eigenvalue}, the plot of \citet{hudson1989source} shown in Fig. \ref{fig:hudson-region} sees more frequent use for historical reasons. The horizontal and vertical coordinates of a moment tensor on this plot are related to its eigenvalues by
\begin{align}
u & =-\frac{2}{3}\frac{1}{\lambda_{\max}}(\lambda_{1}+\lambda_{3}-2\lambda_{2}),\nonumber \\
v & =\frac{1}{3}\frac{1}{\lambda_{\max}}(\lambda_{1}+\lambda_{2}+\lambda_{3}),\label{eq:hudson-uv}
\end{align}
where $\lambda_{\max}=\max(|\lambda_{1}|,|\lambda_{3}|)$ \citep{vavryvcuk2015moment}. For values of $\nu\in[0,0.5]$, $\lambda_{\max}=-\lambda_{3}$ for the four lune points given in eq. \eqref{eq:lune-points}. It follows that the corresponding points on the Hudson plot are
\begin{align}
(u_{D},v_{D}) & =(0,0),\nonumber \\
(u_{K^{-}},v_{K^{-}}) & =\left(\frac{2}{3}\frac{1-2\nu}{1-\nu},-\frac{1}{3}\frac{1+\nu}{1-\nu}\right),\nonumber \\
(u_{1^{-}},v_{1^{-}}) & =\left(-\frac{4}{3}\frac{1-2\nu}{1-\nu},-\frac{1}{3}\frac{1+\nu}{1-\nu}\right),\nonumber \\
(u_{2^{-}},v_{2^{-}}) & =\left(-\frac{2}{3}\left[1-2\nu\right],-\frac{2}{3}\left[1+\nu\right]\right).
\end{align}
Using eq. \eqref{eq:hudson-uv}, it can be shown that a great-circle arc $\mathbf{n}\cdot\boldsymbol{\Lambda}=0$ on the lune for some $\mathbf{n}=(a,b,c)$ become straight-line segments $h(u,v)=0$ on the Hudson plot, where

\begin{equation}
h(u,v)=\begin{cases}
v-\frac{1}{2(b+2c)}\left[(c-b)u+2(c-a)\right] & \mathrm{if}\,v>\frac{1}{4}u,\\
v-\frac{1}{2(b+2a)}\left[(a-b)u+2(c-a)\right] & \mathrm{otherwise}.
\end{cases}
\end{equation}
The line $v=u/4$ is is dotted in Fig. \ref{fig:Region-on-the} and corresponds to $\lambda_{1}=-\lambda_{3}$. For the segments of interest, $\lambda_{1}\leq-\lambda_{3}$ (that is, $v\leq u/4$), meaning that the arcs defined by eq. \eqref{eq:cdc-normals} become
\begin{align}
h_{1D}(u,v) & =v-\frac{\nu+1}{4(1-2\nu)}u=0,\nonumber \\
\text{\ensuremath{h_{DK}}}(u,v) & =v+\frac{\nu+1}{2(1-2\nu)}u=0,\nonumber \\
h_{2^{-}K^{-}}(u,v) & =v-\frac{\nu+1}{2(2-\nu)}u+\frac{\nu+1}{2-\nu}=0.\label{eq:hudson-lines}
\end{align}
It follows that a moment tensor $\mathbf{M}$ will have a closing-CDC decomposition if and only if its corresponding point $(u,v)$ on the Hudson plot satisfies $h_{1D}(u,v)\leq0$, $\text{\ensuremath{h_{DK}}}(u,v)\leq0$, and $h_{2^{-}K^{-}}(u,v)\geq0$.

\section{Decomposition approach}

\label{sec:Decomposition-approach}

As previously noted by \citet{tape2013classical}, a moment tensor falling within the region described in Section \ref{sec:Bounds} will have infinite closing-CDC decompositions in general. That this is the case is not surprising given that (assuming a fixed $\nu$) describing the closing-crack and DC tensors individually requires a total of seven parameters ($M_{K}$, $\phi$, $\beta$ for $\mathbf{M}^{K^{-}}$ and $M_{D}$, $\phi_{s}$, $\delta$, $\lambda$ for $\mathbf{M}^{D}$), while their sum $\mathbf{M}$ requires only six. In this section, we provide a method for generating a subset of the infinite possible decompositions and propose physically motivated approaches for selecting a decomposition from this set.

\subsection{Decomposition generation}

\label{subsec:Parameterisation}

Supposing that we can decompose a given moment tensor $\mathbf{M}$ according to eq. \eqref{eq:source-model}, we can constrain the parameters of the closing-crack component $\mathbf{M}^{K^{-}}(M_{K},\phi,\beta)$ in terms of invariants of the remaining DC component $\mathbf{M}^{D}=\mathbf{M}-\mathbf{M}^{K^{-}}$. The first of these constraints is that $\mathbf{M}^{D}$ must have zero trace, yielding
\begin{align}
\mathrm{tr}(\mathbf{M}^{K^{-}}) & =\mathrm{tr(}\mathbf{M})\nonumber \\
\Rightarrow M_{K} & =-\frac{\mathrm{tr(}\mathbf{M})}{(\nu+1)\alpha_{v}}.\label{eq:crack-scalar-moment}
\end{align}
The second constraint is that the determinant of $\mathbf{M}^{D}$ must also be zero, which defines the implicit equation
\begin{equation}
d(\phi,\beta)=\det\left[\mathbf{M}-\mathbf{M}^{K^{-}}\left(-\frac{\mathrm{tr(}\mathbf{M})}{(\nu+1)\alpha_{v}},\phi,\beta\right)\right]=0\label{eq:det-constraint}
\end{equation}
for the possible $P$-axis orientations of $\mathbf{M}^{K^{-}}$.

\begin{figure}
\begin{centering}
\subfloat[\label{fig:crush-p-contour-stereonet}]{\includegraphics[scale=0.35]{ 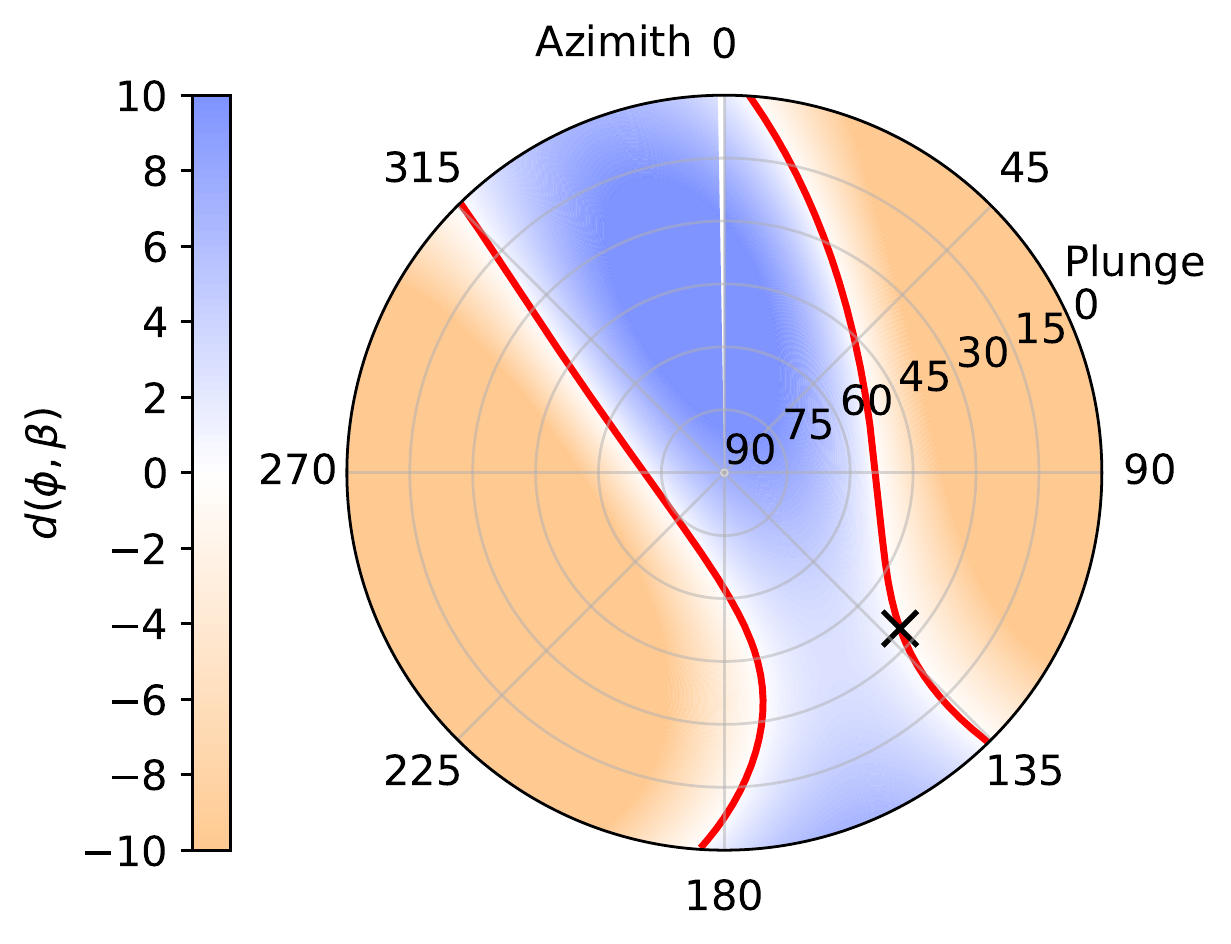}}\subfloat[\label{fig:dc-nodal-pole-steroenet}]{\includegraphics[scale=0.35]{ 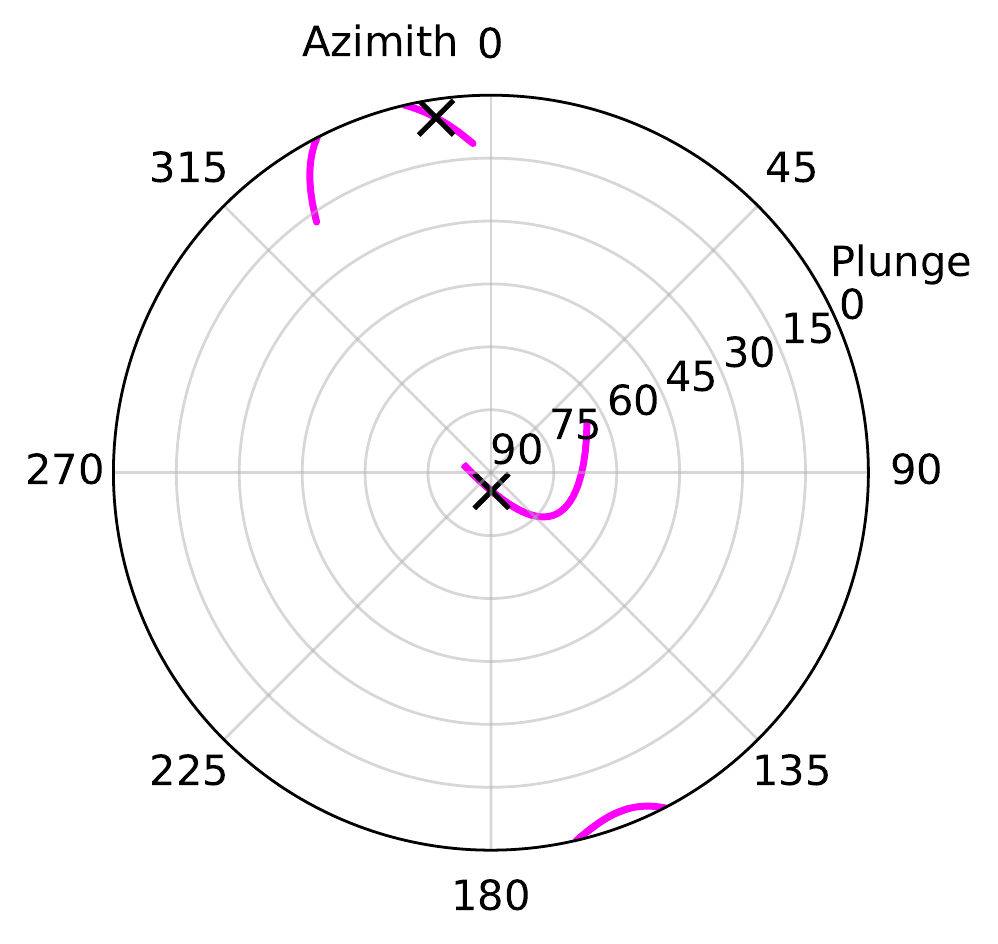}}\vspace{-10bp}
\par\end{centering}
\centering{}\subfloat[\label{fig:beachball-decomp}]{\includegraphics[width=1\columnwidth]{ 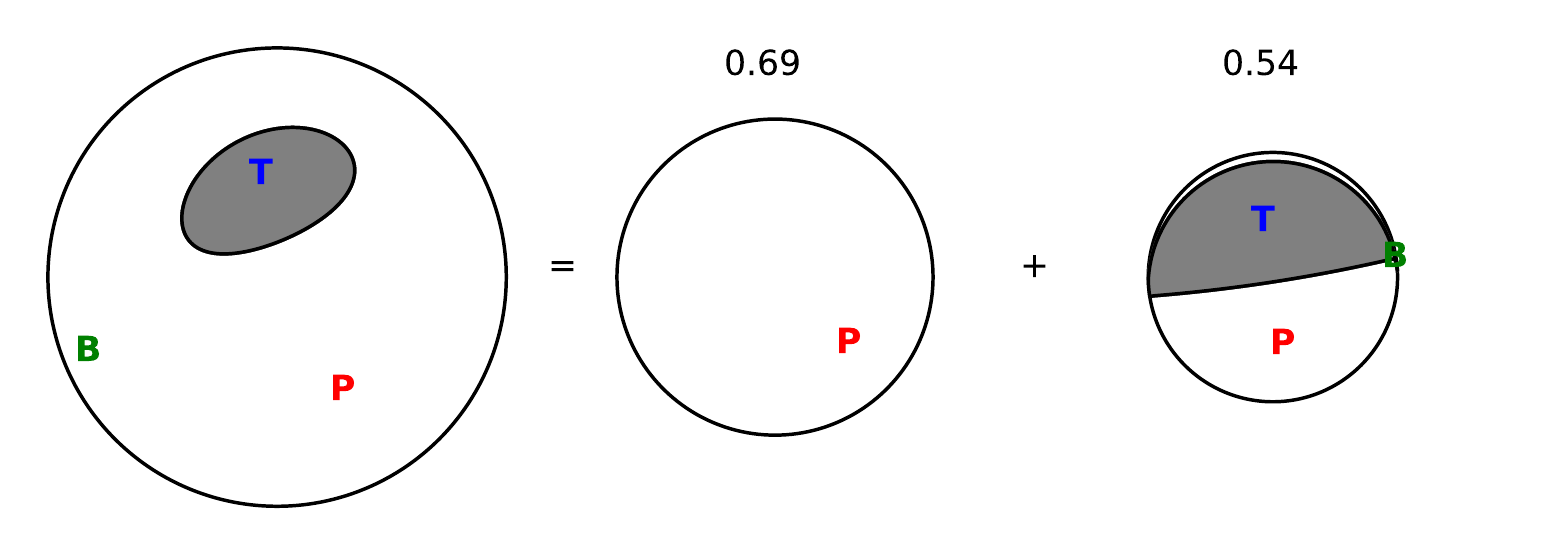}}\caption{(a) Stereonet showing values of $d(\phi,\beta)$ for $\mathbf{M}$ given in eq. \eqref{eq:random-mt}. The $d(\phi,\beta)=0$ contour is shown in red, which corresponds to valid closing-crack $P$-axes. (b) Stereonet of corresponding DC nodal-plane poles. (c) Beachball diagrams for a selected decomposition $\mathbf{M}=\mathbf{M}^{K^{-}}+\mathbf{M}^{D}$ as marked in (a) and (b) with black crosses. The radii of these beachballs is proportional to the listed scalar moment values, which have been normalised to that of $\mathbf{M}$.}
\end{figure}

A simple way of determining a set of closing-crack $P$-axis orientations that approximately satisfy eq. \eqref{eq:det-constraint} is to first calculate $d(\phi,\beta)$ for gridded values of $\phi$ and $\beta$ and then approximate the $d(\phi,\beta)=0$ contour using the marching-squares algorithm \citep{lorensen1987marching}. The results of this procedure for a randomly generated closing-CDC moment tensor
\begin{equation}
\text{\ensuremath{\mathbf{M}}}=\left(\begin{array}{ccc}
-0.56 & 0.21 & -0.71\\
0.21 & -0.52 & 0.28\\
-0.71 & 0.28 & -0.39
\end{array}\right)\label{eq:random-mt}
\end{equation}
are shown on a lower-hemisphere stereographic projection (stereonet) in Fig. \ref{fig:crush-p-contour-stereonet}, with the $d(\phi,\beta)=0$ contour shown in red. For each valid closing-crack $P$-axis orientation, we can determine the nodal planes of the remaining DC component {[}a procedure for doing this is given by \citet{gasperini2003fpspack}{]}, which we present in terms of their poles in Fig. \ref{fig:dc-nodal-pole-steroenet}. Fig. \ref{fig:beachball-decomp} shows the beachball diagrams of a selected decomposition that corresponds to the marked orientations of the closing-crack $P$-axis and nodal-plane poles in Figs. \ref{fig:crush-p-contour-stereonet} and \ref{fig:dc-nodal-pole-steroenet}, respectively. Normalised scalar moments $M_{K}/M$ and $M_{D}/M$ are listed above the closing-crack and DC components, where $M$ is the scalar moment of $\mathbf{M}$. Note that while $M_{K}$ is fixed according to eq. \eqref{eq:crack-scalar-moment}, $M_{D}$ will vary between different decompositions.

\subsection{Decomposition selection}

\label{subsec:Selection}

Depending on the information available about a given event, there are a number of ways in which a single decomposition could be selected:
\begin{itemize}
\item If there is knowledge about the excavation geometry and stress state at/near the source, we can determine the expected $P$-axis orientation of the closing-crack component (see Section \ref{subsec:Crush-type-sources}). The decomposition with a closing-crack $P$-axis closest to this can then be selected (that is, the one minimising the angle to the expected axis).
\item Alternatively, there may be information from geological mapping, seismic data, or some other means available about geological structures at/near the source. This can then be used to determine an expected nodal-plane orientation of the DC component. The decomposition with a DC component that has a nodal plane most similar to this expectation can then be selected. There are a number of ways in which this similarity could be quantified; a simple metric is the angle between the poles.
\item In the absence of information from which the expected orientation of either the closing-crack or DC can be inferred, we can constrain their relative orientations. In particular, we can select the decomposition that for which the two components have the nearest $P$-axis orientations. This is the method used for selecting the decomposition shown in Fig. \ref{fig:beachball-decomp}. Alternatively, if a source is expected to be essentially pure crush- or slip/shear-type, then we could select decomposition whose DC component has the largest or smallest scalar moment $M_{D}$, respectively.
\end{itemize}

\section{Approximate decomposition}

\label{sec:Approximate-decomposition}

A moment tensor $\mathbf{M}$ falling outside the closing-CDC region defined in Section \ref{sec:Bounds} cannot be decomposed directly following the procedure of Section \ref{sec:Decomposition-approach}. However, we can determine and decompose the closest closing-CDC moment tensor $\mathbf{M}^{\mathrm{CDC^{-}}}$ to $\mathbf{M}$, which we take to be the one minimising $\lVert\mathbf{M}-\mathbf{M}^{\mathrm{CDC^{-}}}\rVert$. In other words, $\mathbf{M}^{\mathrm{CDC^{-}}}$ minimises the scalar moment of the remaining non-closing-CDC component $\mathbf{M}^{\mathrm{NCDC^{-}}}=\mathbf{M}-\mathbf{M}^{\mathrm{CDC^{-}}}$. We outline a relatively simple procedure for determining $\mathbf{M}^{\mathrm{CDC^{-}}}$ in this section.

\subsection{Geometry on the lune}

As shown by \citet{tape2012geometric}, $\lVert\mathbf{M}-\mathbf{M}^{\mathrm{CDC^{-}}}\rVert$ is minimised when $\mathbf{M}^{\mathrm{CDC^{-}}}$ has the same eigenframe as $\mathbf{M}$, in which case it equal to the distance $\lVert\boldsymbol{\Lambda}-\boldsymbol{\Lambda}^{\mathrm{CDC^{-}}}\rVert$ between the corresponding eigenvalue triples. This distance satisfies
\begin{align}
\lVert\boldsymbol{\Lambda}-\boldsymbol{\Lambda}^{\mathrm{CDC^{-}}}\rVert^{2} & =\lVert\boldsymbol{\Lambda}\rVert^{2}+\lVert\boldsymbol{\Lambda}^{\mathrm{CDC^{-}}}\rVert^{2}\nonumber \\
 & \phantom{=}-2\lVert\boldsymbol{\Lambda}\rVert\lVert\boldsymbol{\Lambda}^{\mathrm{CDC^{-}}}\rVert\cos\angle(\hat{\boldsymbol{\Lambda}},\hat{\boldsymbol{\Lambda}}^{\mathrm{CDC^{-}}}),\label{eq:eig-distance}
\end{align}
where $\angle(\hat{\boldsymbol{\Lambda}},\hat{\boldsymbol{\Lambda}}^{\mathrm{CDC^{-}}})=\angle(\boldsymbol{\Lambda},\boldsymbol{\Lambda}^{\mathrm{CDC^{-}}})$ is the angle between the triples. For a given angle, eq. \eqref{eq:eig-distance} is minimised for 
\begin{equation}
\lVert\boldsymbol{\Lambda}^{\mathrm{CDC^{-}}}\rVert=\lVert\boldsymbol{\Lambda}\rVert\cos\angle(\hat{\boldsymbol{\Lambda}},\hat{\boldsymbol{\Lambda}}^{\mathrm{CDC^{-}}}),\label{eq:cdc-angle}
\end{equation}
which gives
\begin{equation}
\lVert\boldsymbol{\Lambda}-\boldsymbol{\Lambda}^{\mathrm{CDC^{-}}}\rVert=\lVert\boldsymbol{\Lambda}\rVert\sin\angle(\hat{\boldsymbol{\Lambda}},\hat{\boldsymbol{\Lambda}}^{\mathrm{CDC^{-}}}).\label{eq:ncdc-angle}
\end{equation}
Noting that $\angle(\hat{\boldsymbol{\Lambda}},\hat{\boldsymbol{\Lambda}}^{\mathrm{CDC^{-}}})\in[0,90^{\circ}]$, this means that minimising $\lVert\boldsymbol{\Lambda}-\boldsymbol{\Lambda}^{\mathrm{CDC^{-}}}\rVert$ reduces to finding the closest point $\hat{\boldsymbol{\Lambda}}^{\mathrm{CDC^{-}}}\in\mathbb{L}_{\mathrm{CDC^{-}}}$ to $\hat{\boldsymbol{\Lambda}}$. 

\subsection{Nearest-closing-CDC cases}

\label{subsec:Cases}

\begin{figure}
\begin{centering}
\subfloat[\label{fig:lune-region-1}]{\begin{centering}
\includegraphics[bb=0bp 0bp 224bp 235bp,scale=0.75]{ 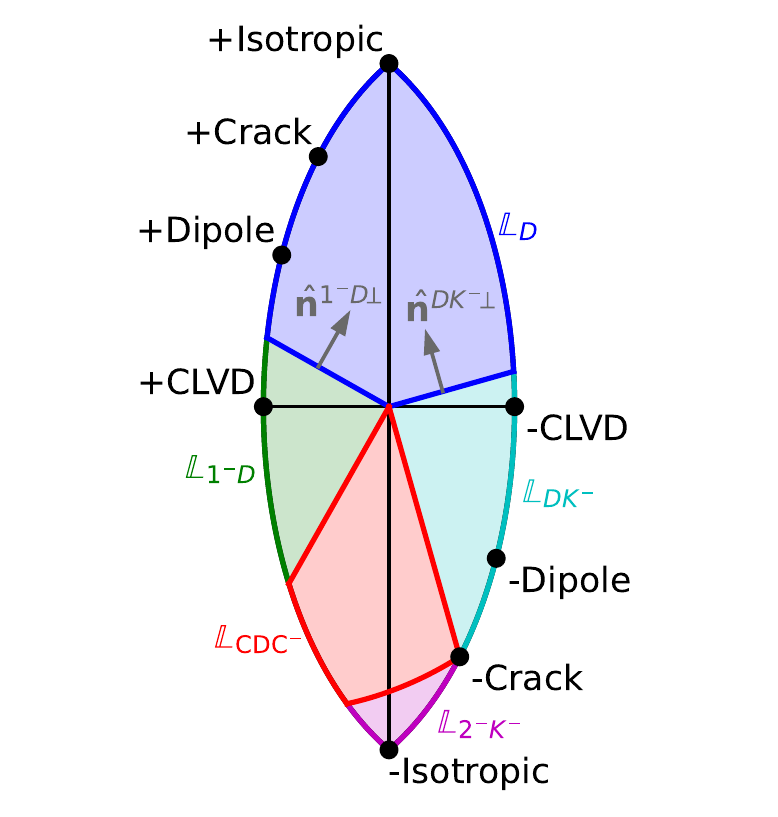}
\par\end{centering}
}
\par\end{centering}
\centering{}\subfloat[\label{fig:hudson-region-1}]{\begin{centering}
\hspace{-25bp}\includegraphics[bb=35bp 0bp 224bp 235bp,scale=0.75]{ 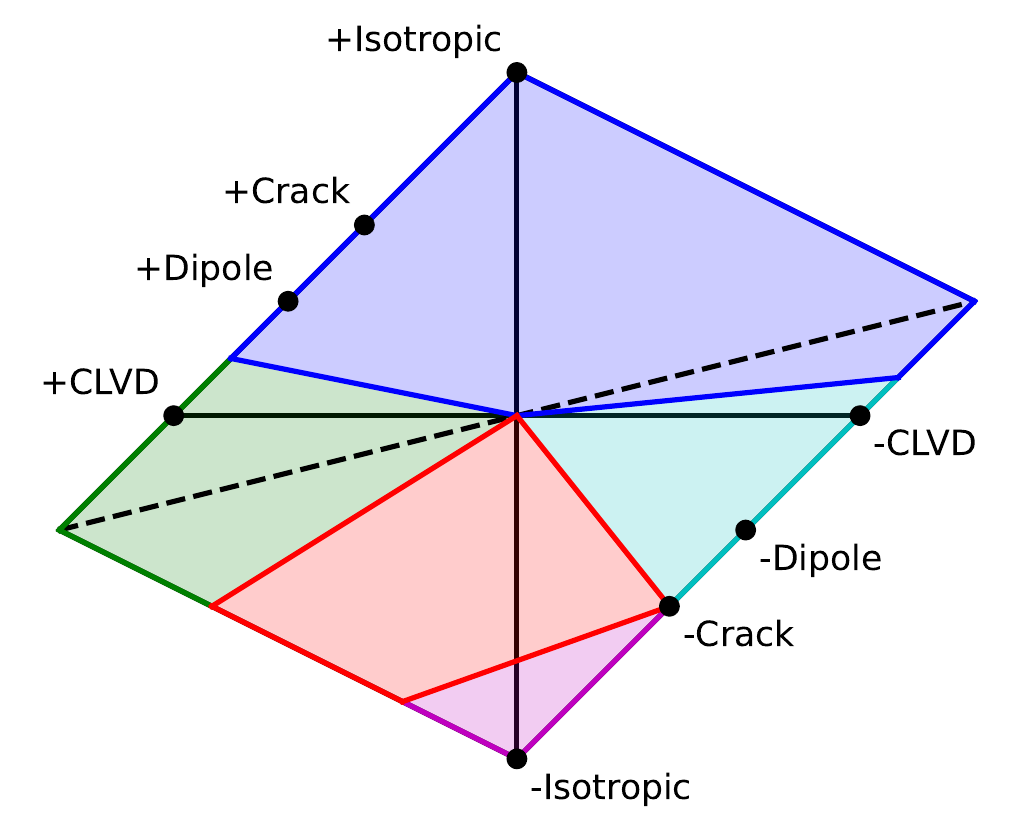}
\par\end{centering}
}\caption{(a) Partition of the lune into the regions defined in eqs. \eqref{eq:cdc-region-def} and \eqref{eq:lune-partition} for $\nu=0.25$. (b) Corresponding regions on the Hudson plot.}
\end{figure}

There are five cases to consider in determining $\hat{\boldsymbol{\Lambda}}^{\mathrm{CDC^{-}}}$, which correspond to the partition of $\mathbb{L}$ into the regions shown in Fig. \ref{fig:lune-region-1}. In addition to $\mathbb{L}_{\mathrm{CDC^{-}}}$ as defined in eq. \eqref{eq:cdc-region-def}, the remaining four regions are
\begin{align}
\mathbb{L}_{D} & =\{\hat{\boldsymbol{\Lambda}}\in\mathbb{L}:\hat{\mathbf{n}}^{1D\perp}\cdot\hat{\boldsymbol{\Lambda}}\geq0,\nonumber \\
 & \phantom{=\{}\hat{\mathbf{n}}^{DK\perp}\cdot\hat{\boldsymbol{\Lambda}}\geq0,\hat{\boldsymbol{\Lambda}}\neq\hat{\boldsymbol{\Lambda}}^{D}\},\nonumber \\
\mathbb{L}_{DK^{-}} & =\{\hat{\boldsymbol{\Lambda}}\in\mathbb{L}:\hat{\mathbf{n}}^{DK\perp}\cdot\hat{\boldsymbol{\Lambda}}<0,\hat{\mathbf{n}}^{DK}\cdot\hat{\boldsymbol{\Lambda}}>0\},\nonumber \\
\mathbb{L}_{1^{-}D} & =\{\hat{\boldsymbol{\Lambda}}\in\mathbb{L}:\hat{\mathbf{n}}^{1D\perp}\cdot\hat{\boldsymbol{\Lambda}}<0,\hat{\mathbf{n}}^{1D}\cdot\hat{\boldsymbol{\Lambda}}>0\},\nonumber \\
\mathbb{L}_{2^{-}K^{-}} & =\{\hat{\boldsymbol{\Lambda}}\in\mathbb{L}:\hat{\mathbf{n}}^{2^{-}K^{-}}\cdot\hat{\boldsymbol{\Lambda}}<0\},\label{eq:lune-partition}
\end{align}
where the normals
\begin{align}
\hat{\mathbf{n}}^{1D\perp} & =p_{\mathbb{S}}(\hat{\mathbf{n}}^{1D}\times\hat{\boldsymbol{\Lambda}}^{D})=p_{\mathbb{S}}(\nu,1-\nu,\nu),\nonumber \\
\hat{\mathbf{n}}^{DK\perp} & =p_{\mathbb{S}}(\hat{\boldsymbol{\Lambda}}^{D}\times\hat{\mathbf{n}}^{DK})=p_{\mathbb{S}}(1,2\nu,1),\label{eq:perp-normals}
\end{align}
define great circles perpendicular at $\hat{\boldsymbol{\Lambda}}^{D}$ to those corresponding to $\hat{\mathbf{n}}^{1D}$ and $\hat{\mathbf{n}}^{DK}$ , respectively. For completeness, we also show the five regions considered on the Hudson plot in Fig. \ref{fig:hudson-region-1}. The straight-line segments corresponding to the normals of eq. \eqref{eq:perp-normals} are
\begin{align}
h_{1D\perp}(u,v) & =v+\frac{1-2\nu}{2(\nu+1)}u=0,\nonumber \\
h_{DK\perp}(u,v) & =v-\frac{1-2\nu}{4(\nu+1)}u=0.
\end{align}

If $\hat{\boldsymbol{\Lambda}}\in\mathbb{L}_{\mathrm{CDC^{-}}}$, then we trivially have $\hat{\boldsymbol{\Lambda}}^{\mathrm{CDC^{-}}}=\hat{\boldsymbol{\Lambda}}$. In the remaining cases, $\hat{\boldsymbol{\Lambda}}^{\mathrm{CDC^{-}}}$ will lie on the boundary of $\mathbb{L}_{\mathrm{CDC^{-}}}$ (see Section \ref{subsec:Lune-source-type-plot} for a description/interpretation of this boundary). The simplest of these cases is when $\hat{\boldsymbol{\Lambda}}\in\mathbb{L}_{D}$, for which $\hat{\boldsymbol{\Lambda}}^{\mathrm{CDC^{-}}}=\hat{\boldsymbol{\Lambda}}^{D}$. If $\hat{\boldsymbol{\Lambda}}\in\mathbb{L}_{DK^{-}}$, then $\hat{\boldsymbol{\Lambda}}^{\mathrm{CDC^{-}}}$ will be the closest point on the great-circle arc joining $\hat{\boldsymbol{\Lambda}}^{D}$ and $\hat{\boldsymbol{\Lambda}}^{K^{-}}$, which corresponds to the plane defined by the normal $\hat{\mathbf{n}}^{DK}$. It is possible to determine $\hat{\boldsymbol{\Lambda}}^{\mathrm{CDC^{-}}}$ by first projecting $\hat{\boldsymbol{\Lambda}}$ onto this plane and then onto the unit sphere; that is, $\hat{\boldsymbol{\Lambda}}^{\mathrm{CDC^{-}}}=p_{\mathbb{S}}(\hat{\boldsymbol{\Lambda}}-[\hat{\boldsymbol{\Lambda}}\cdot\hat{\mathbf{n}}^{DK}]\hat{\mathbf{n}}^{DK})$. The remaining two cases of $\hat{\boldsymbol{\Lambda}}\in\mathbb{L}_{1^{-}D}$ and $\hat{\boldsymbol{\Lambda}}\in\mathbb{L}_{2^{-}K^{-}}$ can be treated in a similar way. In summary, 
\begin{equation}
\hat{\boldsymbol{\Lambda}}^{\mathrm{CDC^{-}}}=\begin{cases}
\hat{\boldsymbol{\Lambda}} & \mathrm{if\,}\hat{\boldsymbol{\Lambda}}\in\mathbb{L}_{\mathrm{CDC^{-}}},\\
\hat{\boldsymbol{\Lambda}}^{D} & \mathrm{if\,}\hat{\boldsymbol{\Lambda}}\in\mathbb{L}_{D},\\
p_{\mathbb{S}}(\hat{\boldsymbol{\Lambda}}-[\hat{\boldsymbol{\Lambda}}\cdot\hat{\mathbf{n}}^{DK}]\hat{\mathbf{n}}^{DK}) & \mathrm{if\,}\hat{\boldsymbol{\Lambda}}\in\mathbb{L}_{DK^{-}},\\
p_{\mathbb{S}}(\hat{\boldsymbol{\Lambda}}-[\hat{\boldsymbol{\Lambda}}\cdot\hat{\mathbf{n}}^{1D}]\hat{\mathbf{n}}^{1D}) & \mathrm{if\,}\hat{\boldsymbol{\Lambda}}\in\mathbb{L}_{1^{-}D},\\
p_{\mathbb{S}}(\hat{\boldsymbol{\Lambda}}-[\hat{\boldsymbol{\Lambda}}\cdot\hat{\mathbf{n}}^{2^{-}K^{-}}]\hat{\mathbf{n}}^{2^{-}K^{-}}) & \mathrm{if\,}\hat{\boldsymbol{\Lambda}}\in\mathbb{L}_{2^{-}K^{-}}.
\end{cases}
\end{equation}

\subsection{Non-closing-CDC content contours}

\label{subsec:Contours}

\begin{figure}
\begin{centering}
\subfloat[\label{fig:lune-isolines}]{\begin{centering}
\includegraphics[bb=0bp 0bp 224bp 235bp,scale=0.75]{ 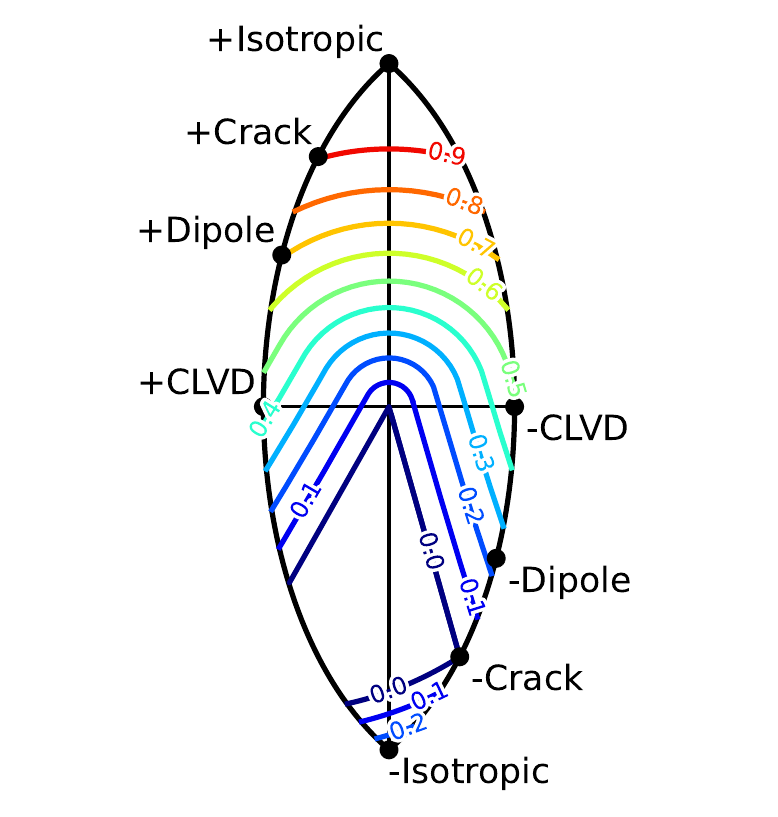}
\par\end{centering}
}
\par\end{centering}
\centering{}\subfloat[\label{fig:hudson-isolines}]{\begin{centering}
\hspace{-25bp}\includegraphics[bb=35bp 0bp 224bp 235bp,scale=0.75]{ 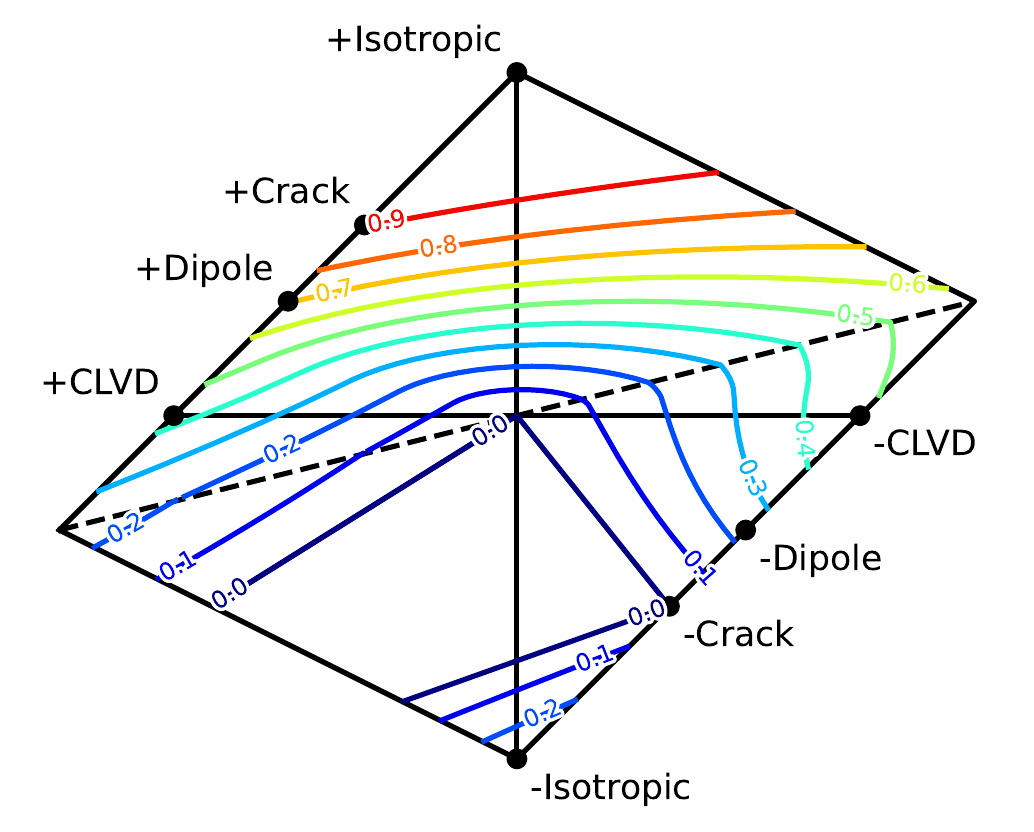}
\par\end{centering}
}\caption{(a) Isolines of constant non-closing-CDC amount as defined by eq. \eqref{eq:non-cdc-amount} on the lune for $\nu=0.25$. (b) Same on the Hudson plot.}
\end{figure}

Rearranging eq. \eqref{eq:ncdc-angle} gives the normalised scalar moment
\begin{equation}
\gamma_{\mathrm{CDC^{-}}}=\frac{\lVert\boldsymbol{\Lambda}-\boldsymbol{\Lambda}^{\mathrm{CDC^{-}}}\rVert}{\lVert\boldsymbol{\Lambda}\rVert}=\sin\angle(\hat{\boldsymbol{\Lambda}},\text{\ensuremath{\hat{\boldsymbol{\Lambda}}^{\mathrm{CDC^{-}}}}}),\label{eq:non-cdc-amount}
\end{equation}
which quantifies the amount of non-closing-CDC content in a given moment tensor. We plot isolines of constant $\gamma_{\mathrm{CDC^{-}}}$ on the lune in Fig. \ref{fig:lune-isolines}, which are simply lines at a constant angle $\sin^{-1}(\gamma_{\mathrm{CDC^{-}}})$ from the boundary of $\mathbb{L}_{\mathrm{CDC^{-}}}$. These lines are composed of small circle sections in the regions defined by eq. \eqref{eq:lune-partition}. In $\mathbb{L}_{D}$, these small circles are at an angle $\sin^{-1}(\gamma_{\mathrm{CDC^{-}}})$ from $\hat{\boldsymbol{\Lambda}}^{D}$. In the remaining regions $\mathbb{L}_{DK^{-}}$, $\mathbb{L}_{1^{-}D}$, and $\mathbb{L}_{2^{-}K^{-}}$, they are at a constant angle of $90^{\circ}-\sin^{-1}(\gamma_{\mathrm{CDC^{-}}})$ from the normals $\hat{\mathbf{n}}^{DK}$, $\hat{\mathbf{n}}^{1D}$, and $-\hat{\mathbf{n}}^{2^{-}K^{-}}$, respectively. Alternatively, we can express an isoline in terms of plane intersections as
\begin{equation}
\begin{cases}
\hat{\boldsymbol{\Lambda}}\cdot\hat{\boldsymbol{\Lambda}}^{D}-\sqrt{1-\gamma_{\mathrm{CDC^{-}}}^{2}}=0 & \mathrm{if}\,\hat{\boldsymbol{\Lambda}}\in\mathbb{L}_{D},\\
\hat{\boldsymbol{\Lambda}}\cdot\hat{\mathbf{n}}^{DK}-\gamma_{\mathrm{CDC^{-}}}=0 & \mathrm{if}\,\hat{\boldsymbol{\Lambda}}\in\mathbb{L}_{DK^{-}},\\
\hat{\boldsymbol{\Lambda}}\cdot\hat{\mathbf{n}}^{1D}-\gamma_{\mathrm{CDC^{-}}}=0 & \mathrm{if}\,\hat{\boldsymbol{\Lambda}}\in\mathbb{L}_{1^{-}D},\\
\hat{\boldsymbol{\Lambda}}\cdot\hat{\mathbf{n}}^{2^{-}K^{-}}+\gamma_{\mathrm{CDC^{-}}}=0 & \mathrm{if}\,\hat{\boldsymbol{\Lambda}}\in\mathbb{L}_{2^{-}K^{-}}.
\end{cases}
\end{equation}
Fig. \ref{fig:hudson-isolines} shows these isolines translated to the Hudson plot using eq. \eqref{eq:hudson-uv}. They are significantly more complicated than those on the lune as small circles do not map to straight-line segments like great circles do.

\section{Case studies}

\label{sec:Example-applications}

The methods outlined in Sections \ref{sec:Decomposition-approach} and \ref{sec:Approximate-decomposition} provide us with a relatively simple means of determining a closing-CDC decomposition for a given moment tensor. In this section, we demonstrate these methods on moment tensors inferred from real seismic events recorded at two mines. 

\subsection{Case A}

\label{subsec:Case-A}

\begin{figure*}
\begin{minipage}[t]{0.5\textwidth}%
\subfloat[\label{fig:mpo-330}]{\includegraphics[bb=0bp 0bp 1080bp 720bp,clip,width=1\textwidth]{ 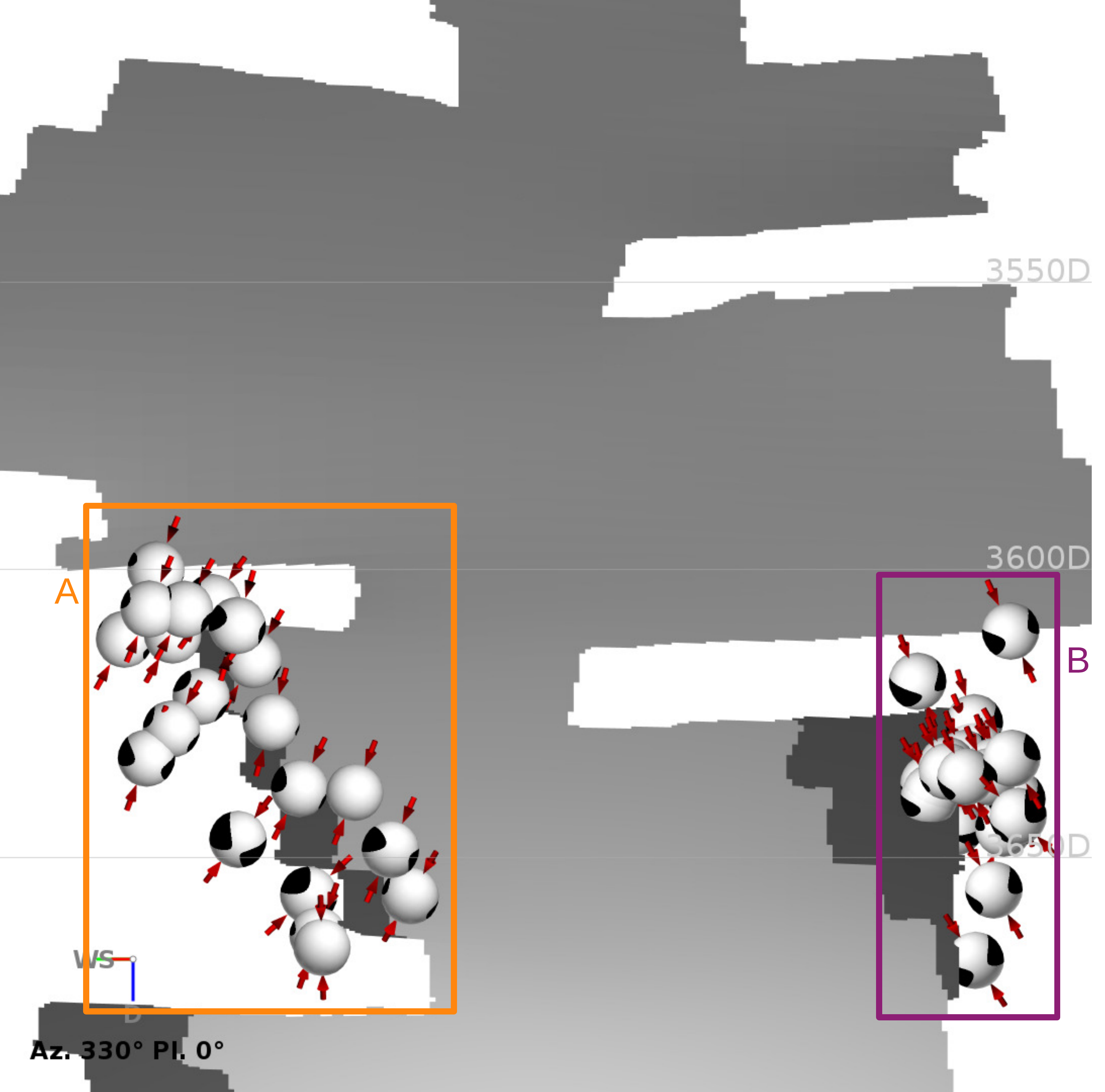}}
\begin{flushleft}
\subfloat[\label{fig:mpo-plan}]{\includegraphics[width=1\textwidth]{ 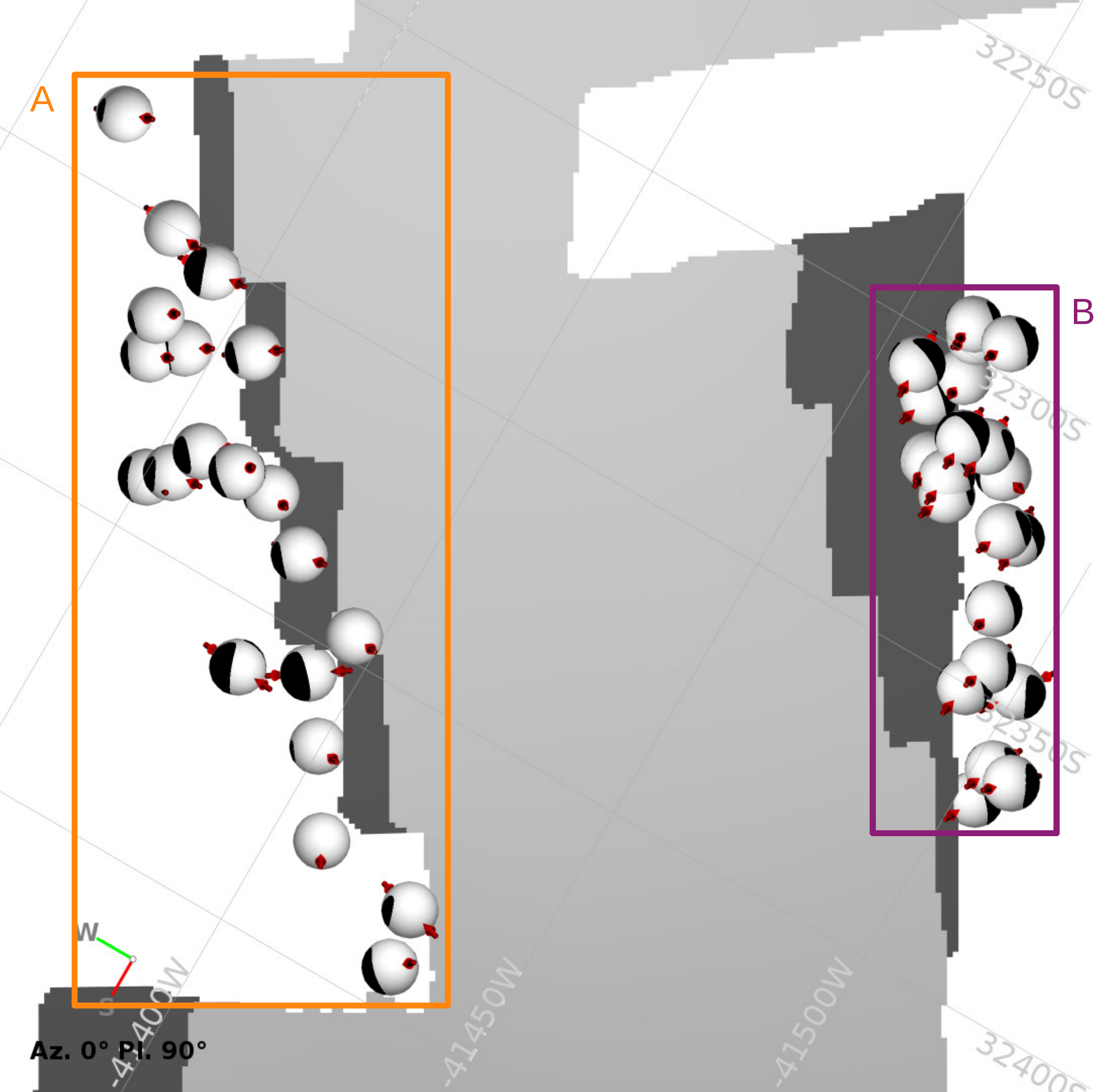}}
\par\end{flushleft}%
\end{minipage}%
\begin{minipage}[t]{0.5\textwidth}%
\begin{flushleft}
\subfloat[\label{fig:mpo-240}]{\includegraphics[bb=0bp 0bp 1080bp 720bp,clip,width=1\textwidth]{ 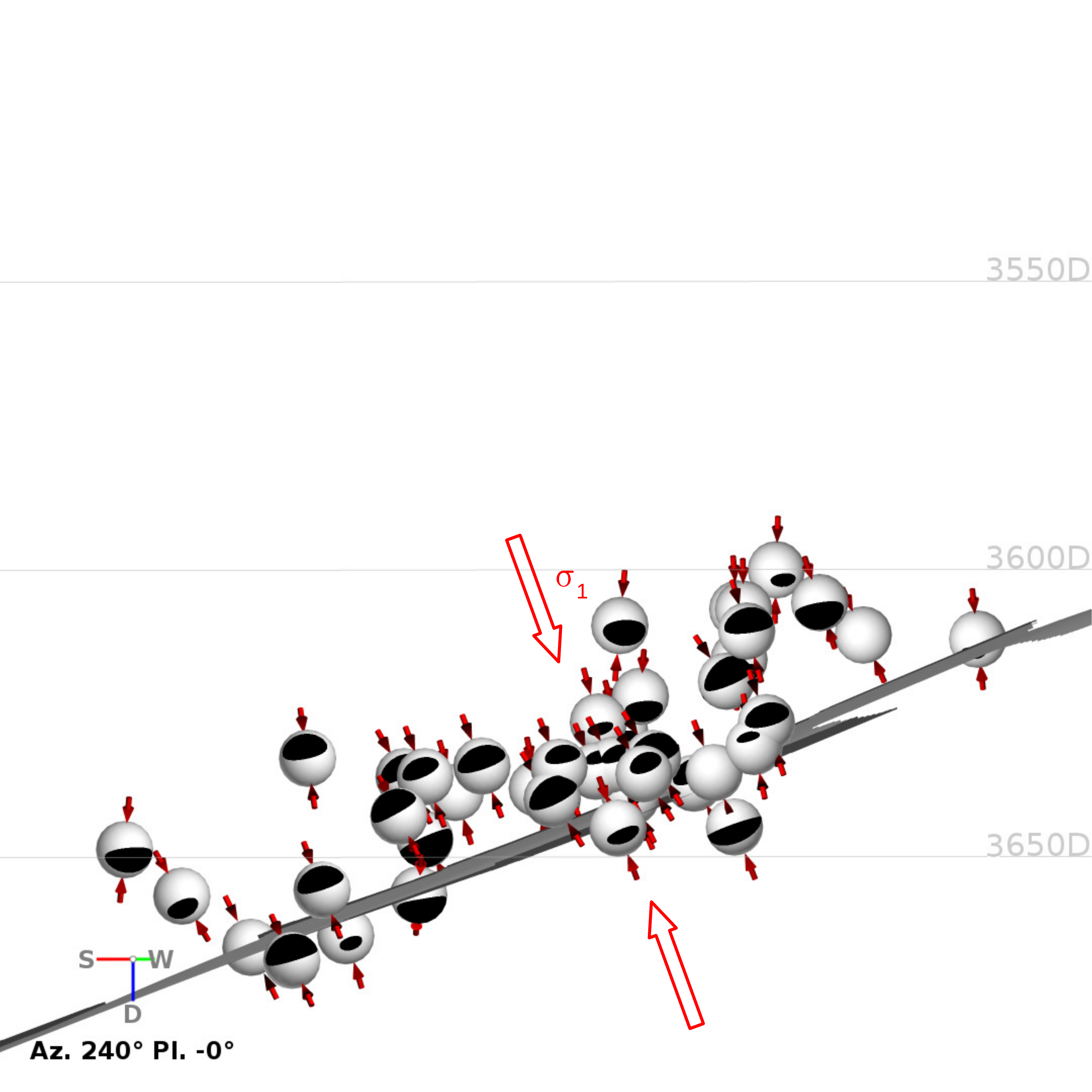}}\vspace{0.75cm}
\par\end{flushleft}
\begin{flushleft}
\subfloat[\label{fig:mpo-hudson}]{\includegraphics[bb=0bp 0bp 298bp 239bp,width=1\textwidth]{ 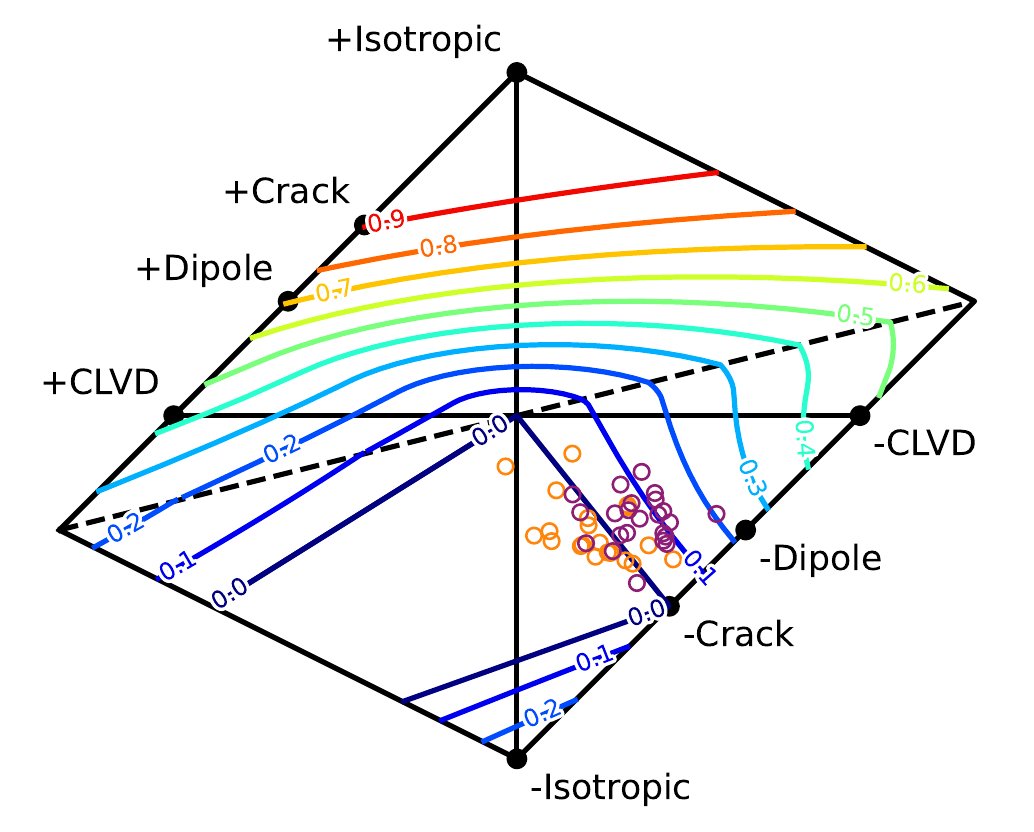}

}
\par\end{flushleft}%
\end{minipage}

\vspace{0.5cm}

\noindent\begin{minipage}[t]{1\textwidth}%
\subfloat[\label{fig:mpo-p}]{~~~\includegraphics[bb=0bp 0bp 298bp 239bp,width=0.33\textwidth]{ 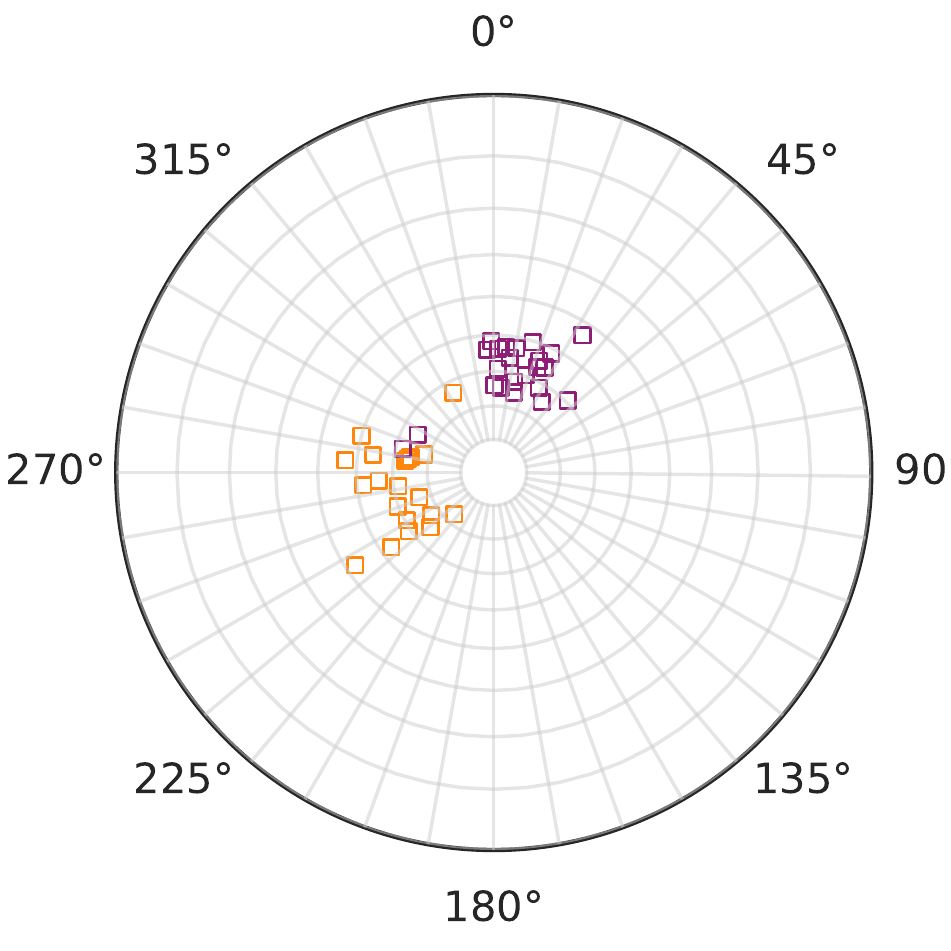}}\subfloat[]{~~~\includegraphics[bb=0bp 0bp 298bp 239bp,width=0.33\textwidth]{ 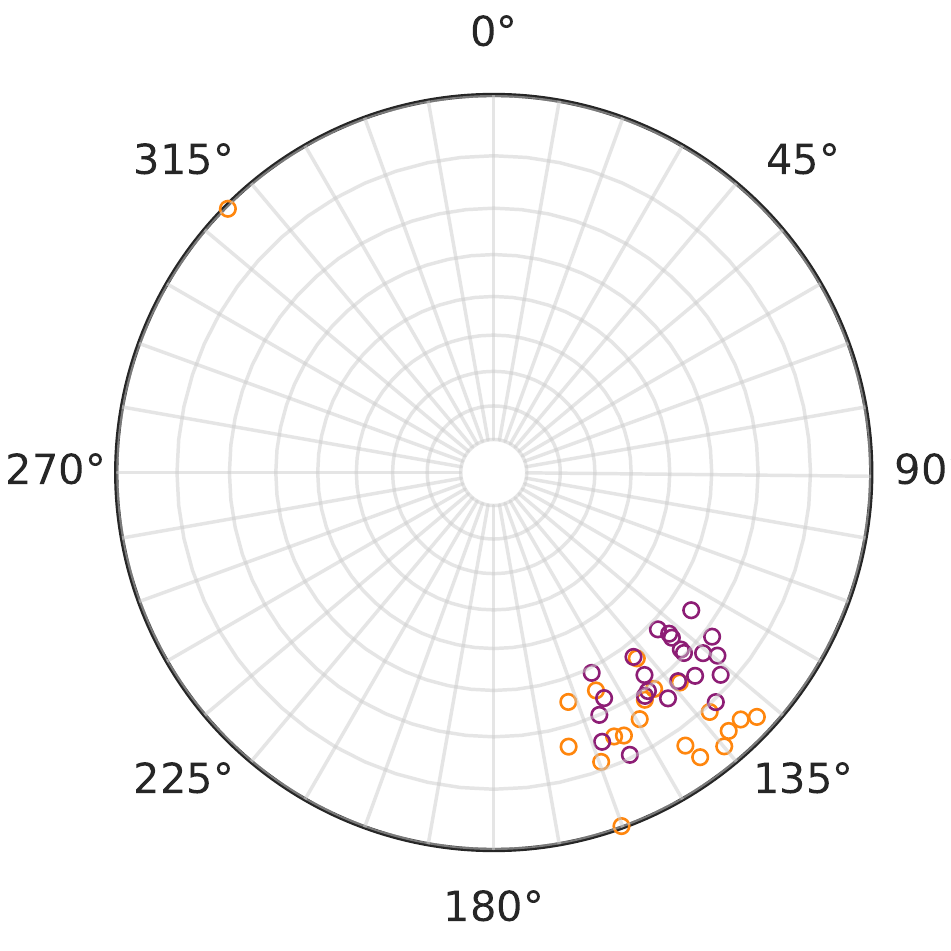}}\subfloat[\label{fig:mpo-t}]{~~~\includegraphics[bb=0bp 0bp 298bp 239bp,width=0.33\textwidth]{ 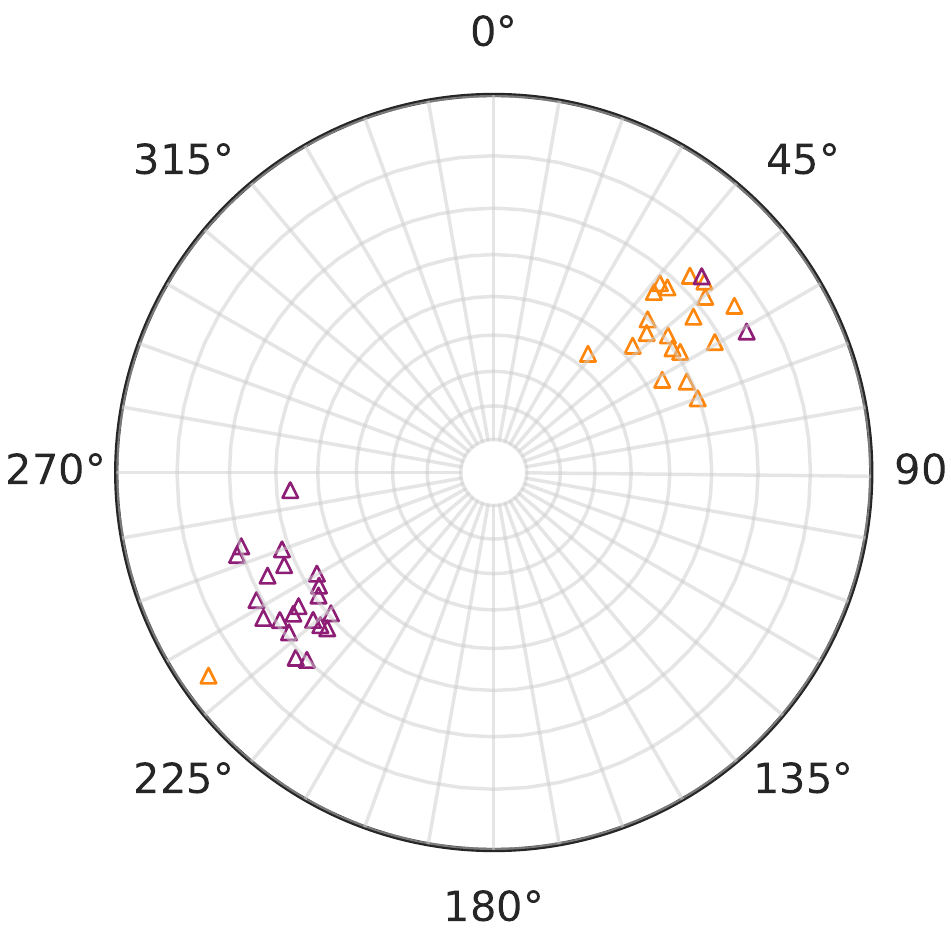}}%
\end{minipage}

\caption{Case study A data summary. (a) View looking at an azimuth of $330^{\circ}$. The footprint of the stope is shown in grey, with darker areas being mined in the period of interest. Seismic events are rendered as beachballs centered at their inferred hypocenter, with red dipoles indicating the $P$-axis direction. Events are grouped spatially into clusters A and B. (b) Plan view (rotated to have an azimuth of $330^{\circ}$ at the top). (c) View at an azimuth of $240^{\circ}$, with the direction of \emph{in situ} maximum a principal compressive stress $\sigma_{1}$ shown. (d) Hudson source-type plot with colouring according to cluster. The contours indicate amount of non-closing-CDC content as defined in Sec \ref{subsec:Contours}. (e)-(g) Stereonet plots showing the $P$-, $B$-, and $T$-axis orientation, respectively, with colouring according to cluster.\label{fig:mpo-summary} }
\end{figure*}

The first catalogue we consider comes from a South African mine and is summarised in Fig. \ref{fig:mpo-summary}. It is composed of 43 events with moment magnitude $M_{W}\geq1.0$ that were recorded in the first half of 2022 near the two advancing fronts of a tabular stope. This stope, the footprint (shape in the plane of the orebody) of which is shown in Figs. \ref{fig:mpo-330}-\ref{fig:mpo-240}, has a height of approximately $\unit[1]{m}$, strike of $60^{\circ}$, and dip of $22^{\circ}$. Also shown is the spatial distribution of the events, which are rendered as three-dimensional beachballs. We have labeled the spatial clusters corresponding to the south-western and north-eastern fronts as A and B, respectively. The spread of event locations in the direction normal to the stope is mostly an artefact of the configuration of the mine's seismic array, which lies largely in the plane of the stope. As shown in Fig. \ref{fig:mpo-hudson}, the moment tensors for events in both clusters lie either in or near the closing-CDC region (for $\nu=0.25$) described in Section \ref{sec:Bounds}. Their orientations are reasonably consistent within each cluster as shown in Figs. \ref{fig:mpo-p}-\ref{fig:mpo-t}. We note that the uncertainty in event locations is not expected to have a significant impact on the inferred moment tensors. This is because the spread in locations in the direction normal to the stope corresponds to a variation in source-station angle of less than $5^{\circ}$ for the stations used in inversion.

To aid in their interpretation, we have determined a closing-CDC decomposition for each moment tensor $\mathbf{M}$ in three steps: First, the closest closing-CDC moment tensor $\mathbf{M}^{\mathrm{CDC^{-}}}$ to $\mathbf{M}$ is found following the procedure of Section \ref{subsec:Cases}. A set of possible decompositions of $\mathbf{M}^{\mathrm{CDC^{-}}}$ is then determined as outlined in Section \ref{subsec:Parameterisation}. Lastly, we select a single decomposition from this set using the first criteria given in Section \ref{subsec:Selection}; in particular, we choose the decomposition for which the closing-crack component's $P$-axis is as close as possible to being orthogonal to the plane of the stope, which has a normal vector with an azimuth of $330^{\circ}$ and plunge of $68^{\circ}$.

\begin{figure}
\subfloat[\label{fig:cluster-A-crush-p}]{\includegraphics[width=0.48\columnwidth]{ 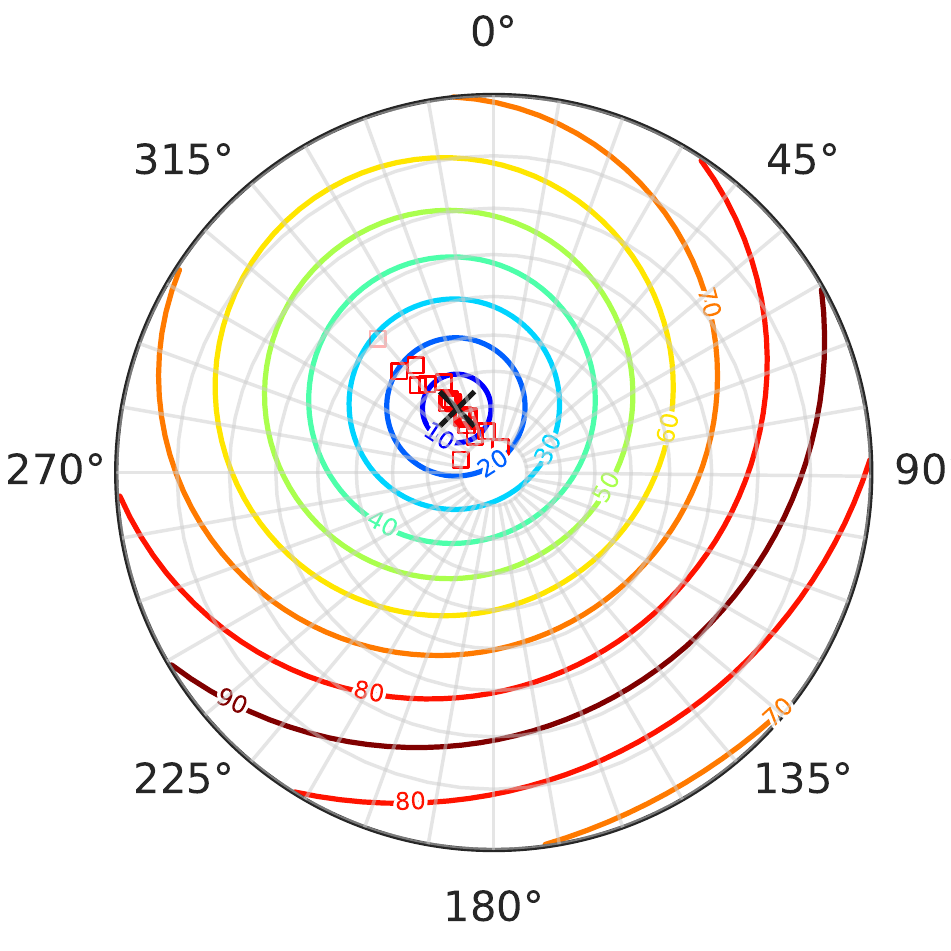}}\subfloat[\label{fig:cluster-A-dc-nodal}]{\includegraphics[width=0.48\columnwidth]{ 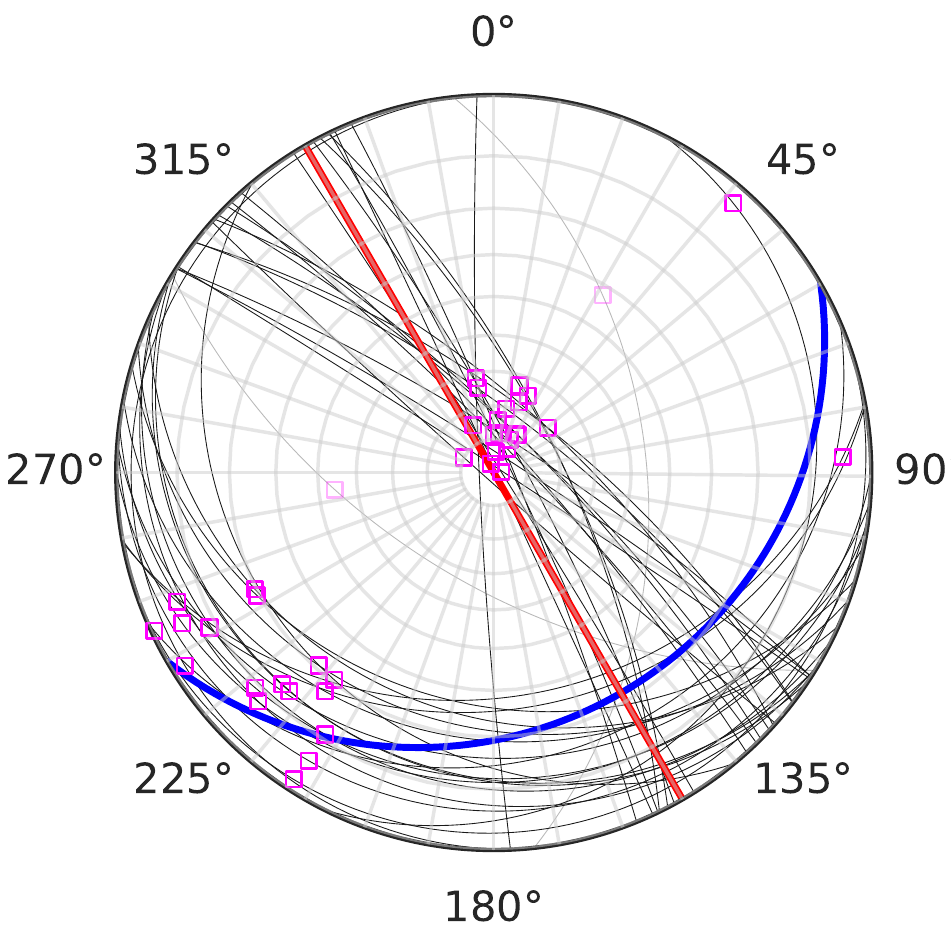}}\vspace{-0.3cm}

\subfloat[\label{fig:cluster-B-crush-p}]{\includegraphics[width=0.48\columnwidth]{ 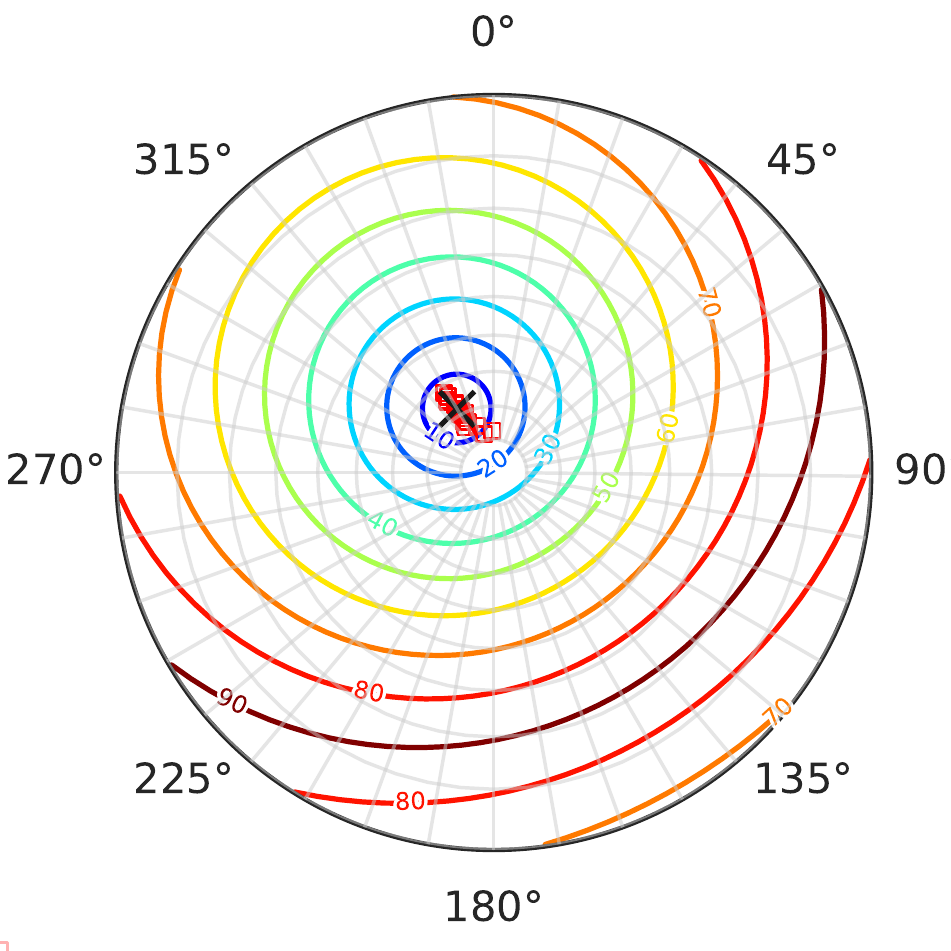}}\subfloat[\label{fig:cluster-B-dc-nodal}]{\includegraphics[width=0.48\columnwidth]{ 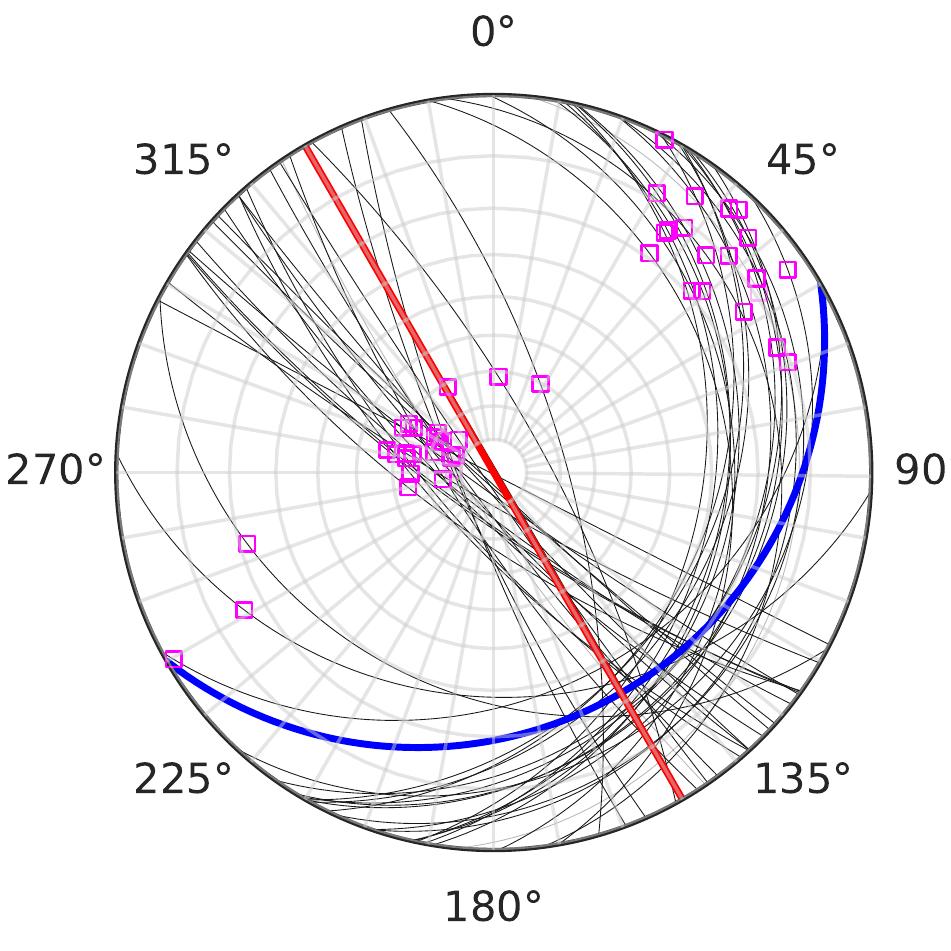}}\vspace{-0.3cm}

\subfloat[\label{fig:mpo-scalar-moment-ratios}]{\includegraphics[width=0.96\columnwidth]{ 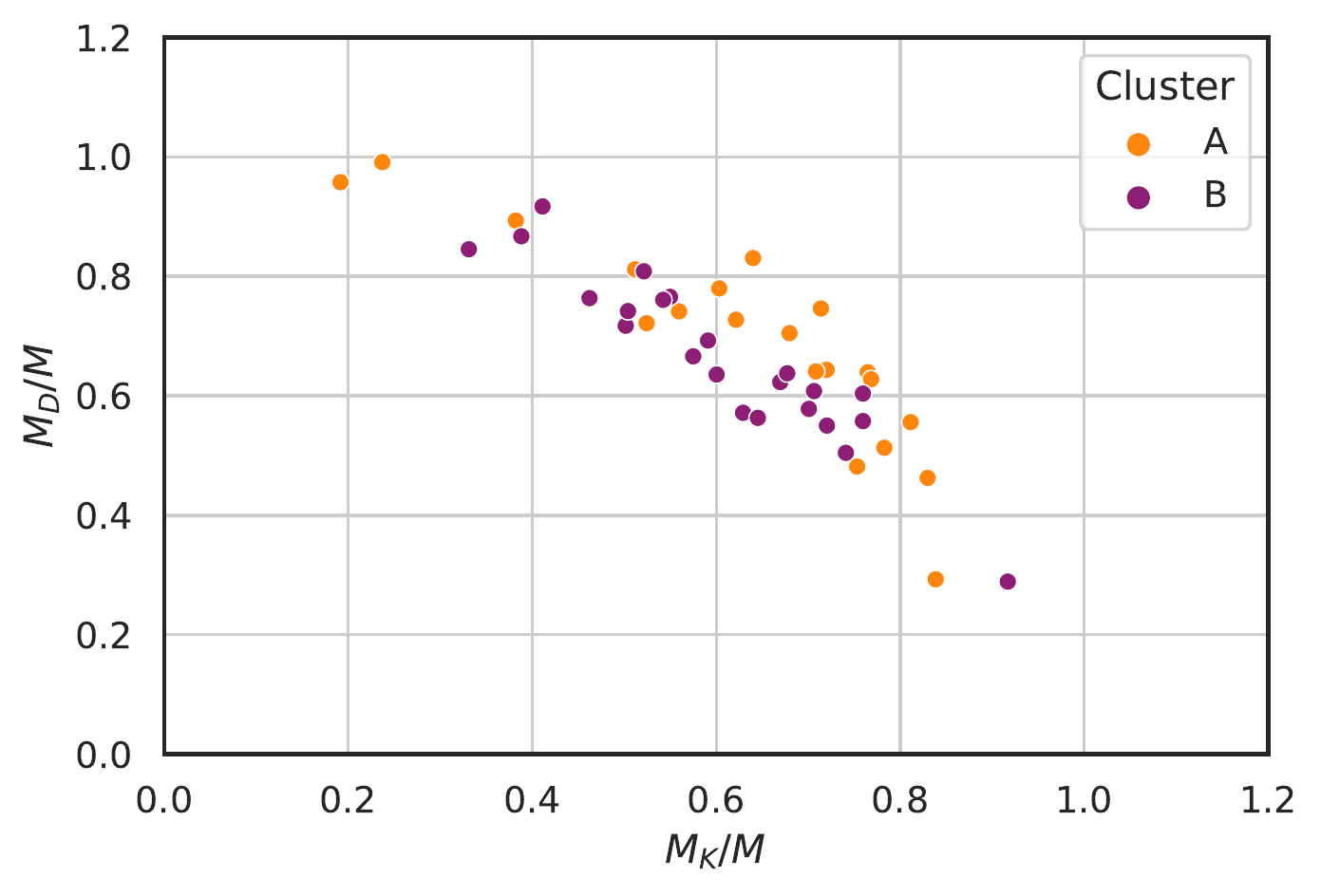}

}

\caption{Case study A decomposition summary. (a) Stereonet of closing-crack $P$-axis orientations for cluster A (red squares). Those that are translucent correspond to decompositions with a small closing-crack component ($M_{K}/M<0.3$). The black cross gives the direction orthogonal to the plane of the stope. The contours give angle to this direction. (b) Stereonet of the DC nodal planes (black lines) and poles (magenta squares) for cluster A. Those that are translucent correspond to decompositions with a small DC component ($M_{D}/M<0.3$). The blue and red lines give the orientation of the stope and its face, respectively. (c) Same as (a) for cluster B. (d) Same as (b) for cluster B. (e) Normalised scalar moments for the closing-crack and DC components coloured by cluster.\label{fig:(a)-Stereonet-of}}
\end{figure}

The closing-crack $P$-axis orientations for cluster A are shown relative to the stope's normal vector in Fig. \ref{fig:cluster-A-crush-p}. The greatest deviation from this normal is by approximately $30^{\circ}$. However, this is for a moment tensor whose decomposition has a small closing-crack component; in particular, it has $M_{K}/M=0.24$. As such, it is reasonable to interpret the source of this event as being a near-pure episode of slip/shear, with the closing-crack component being poorly constrained and possibly just an artefact of the inversion/decomposition procedure. We have rendered the $P$-axes for all decompositions with $M_{K}/M<0.3$ translucently to indicate that they should be interpreted with caution. It can be seen in Fig. \ref{fig:mpo-scalar-moment-ratios} that there are two such decompositions. 

The nodal planes of the DC components for cluster A are shown in Fig. \ref{fig:cluster-A-dc-nodal}. Here, the sole decomposition with $M_{D}/M<0.3$ is rendered transparently and is reasonably interpreted as near-pure crush-type source. For the majority of the remaining decompositions, it can be seen that they have a DC nodal plane that dips towards the face of the stope (to the north east), which is shown in red. These can be interpreted as shearing ahead of the face in the hanging wall as shown in Fig. \ref{fig:Section-view-of-1}. Alternatively, these sources could be interpreted as episodes of slip/shear in the plane of the stope, which is shown in blue in Fig. \ref{fig:cluster-A-dc-nodal}; however, this is a less likely mode of failure given that the direction of maximum loading is approximately orthogonal to the plane of the stope as shown in Fig. \ref{fig:mpo-240}.

The closing-crack $P$-axis orientations for cluster B are shown in Fig. \ref{fig:cluster-B-crush-p}, which are all within $15^{\circ}$ of the stope's normal. Again, as shown in Fig. \ref{fig:cluster-B-dc-nodal}, the majority of the DC components have a nodal plane dipping towards the face of the stope (to the south west), which leads to an interpretation of shearing in the hanging wall ahead of the face. However, there are also two that have nodal planes dipping away from the face of the stope, which instead points towards shearing in the footwall. 

\subsection{Case B}

\begin{figure*}
\begin{minipage}[t]{0.5\textwidth}%
\subfloat[\label{fig:atn-west}]{\includegraphics[width=1\textwidth]{ 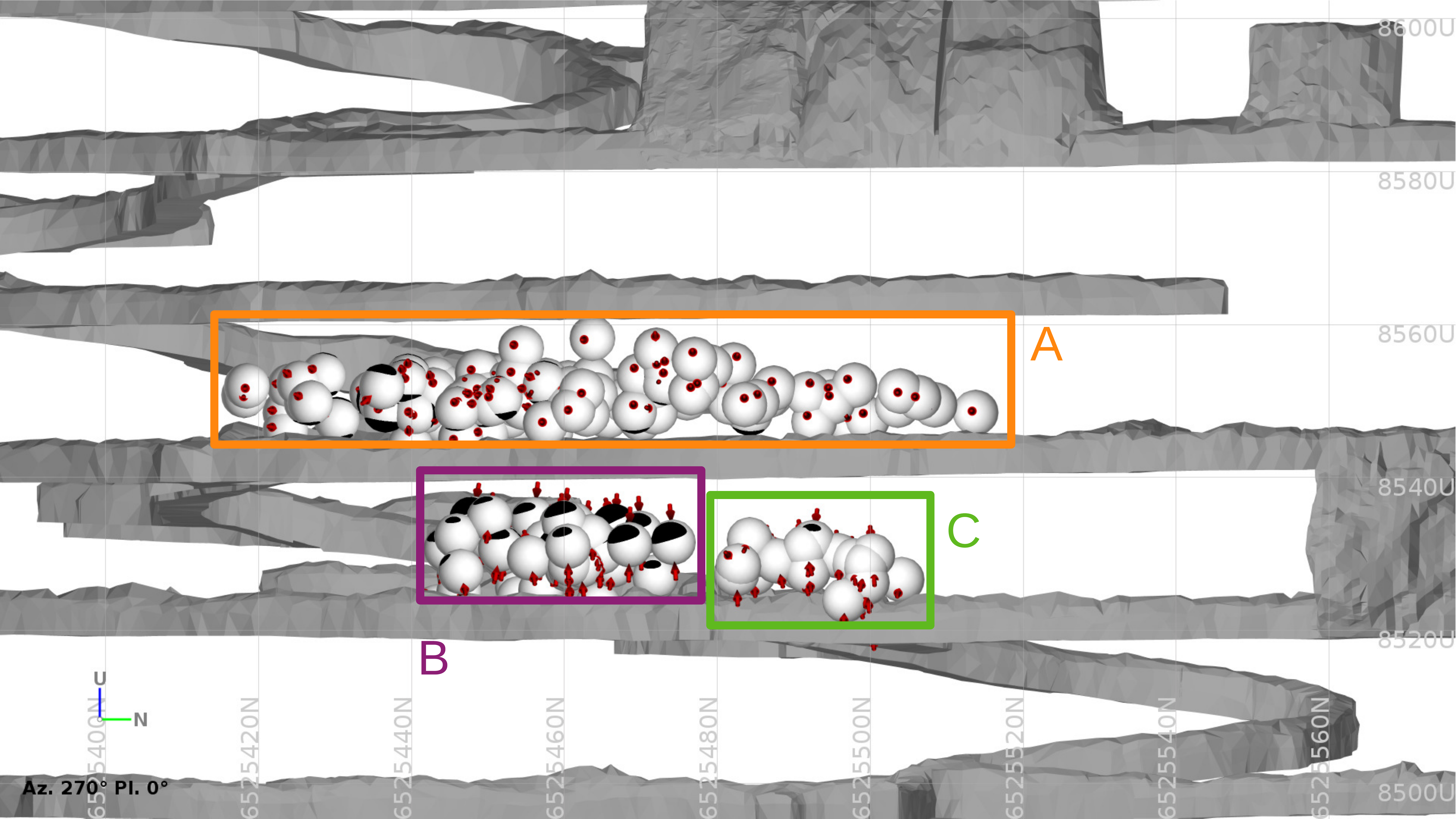}}
\begin{flushleft}
\subfloat[\label{fig:atn-plan-upper}]{\includegraphics[bb=0bp 0bp 1920bp 850bp,clip,width=1\textwidth]{ 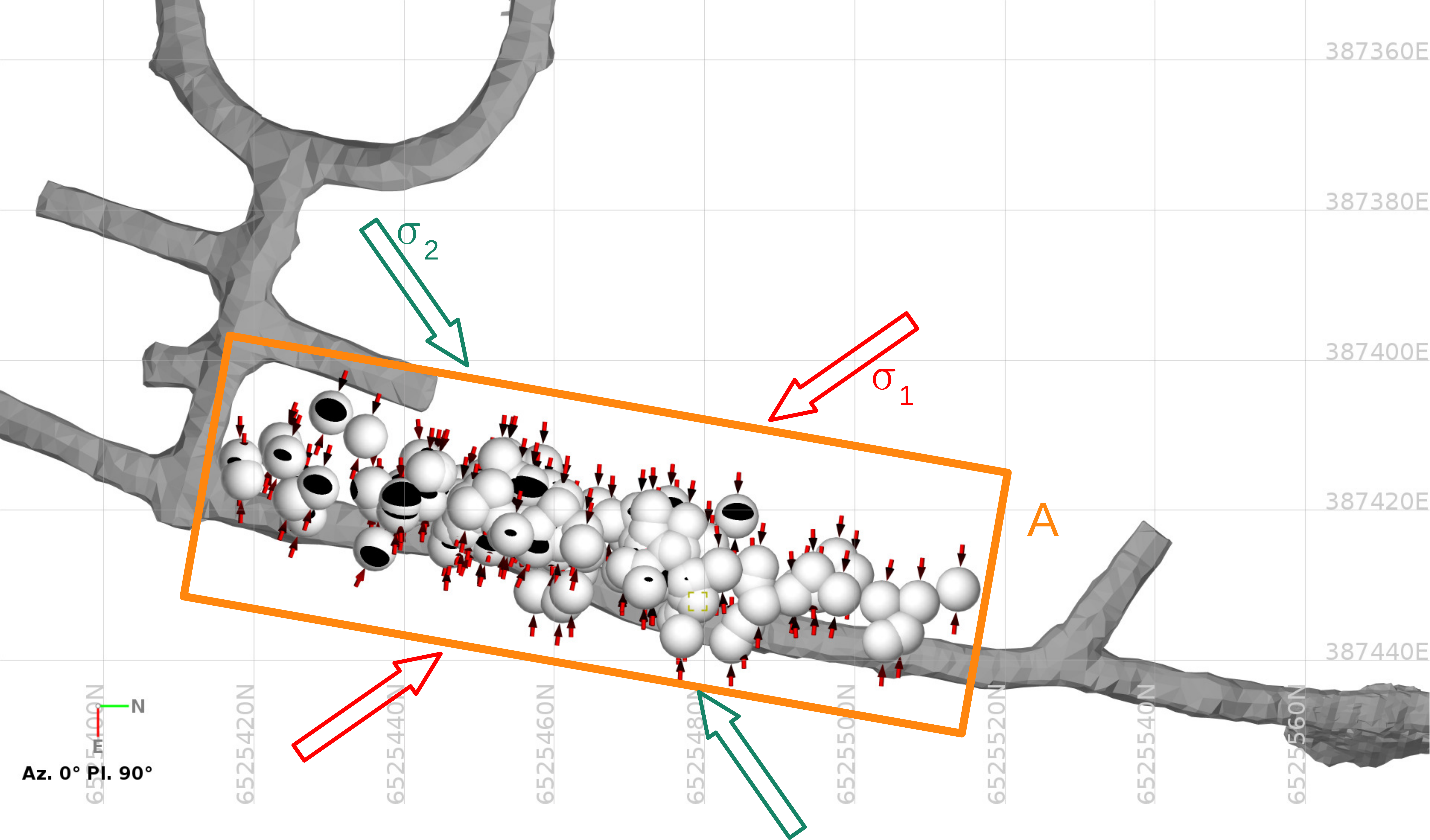}}
\par\end{flushleft}
\begin{flushleft}
\subfloat[\label{fig:atn-plan-lower}]{\includegraphics[bb=0bp 0bp 1920bp 850bp,clip,width=1\textwidth]{ 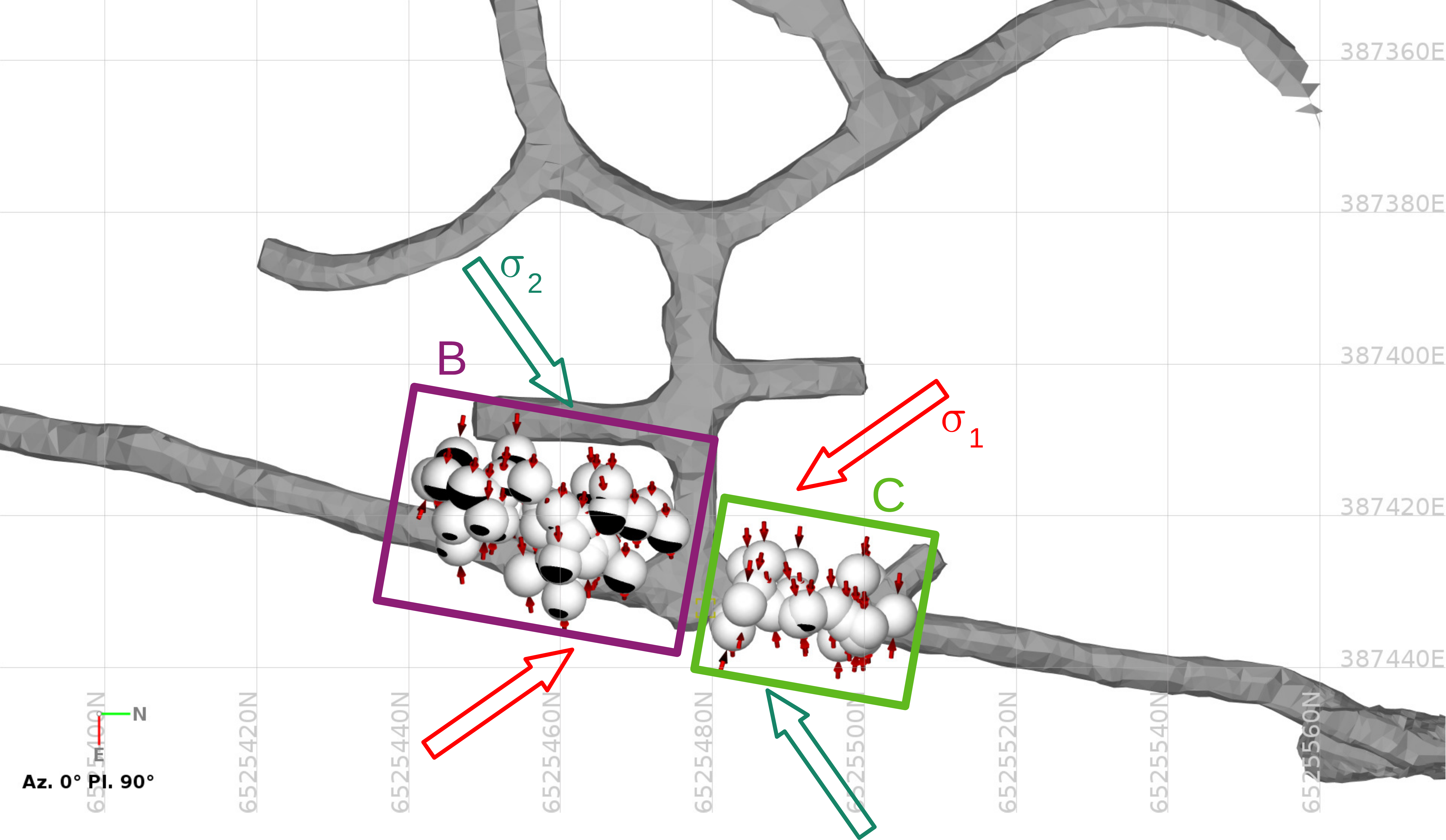}}
\par\end{flushleft}%
\end{minipage}%
\begin{minipage}[t]{0.5\textwidth}%
\subfloat[\label{fig:atn-south}]{\includegraphics[width=1\textwidth]{ 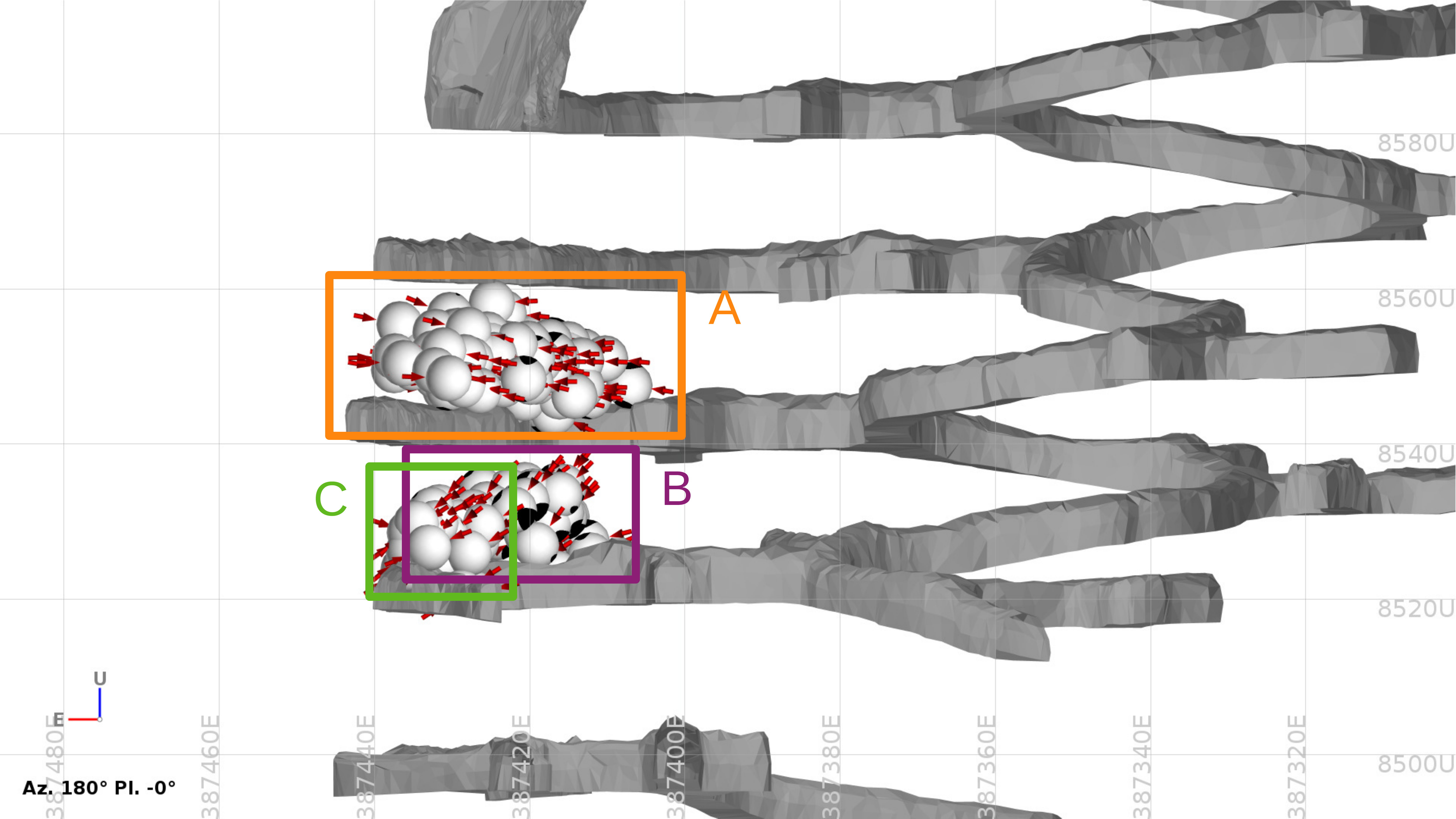}}\vspace{0.75cm}

\begin{flushleft}
\subfloat[\label{fig:atn-hudson}]{\includegraphics[bb=0bp 0bp 298bp 239bp,width=1\textwidth]{ 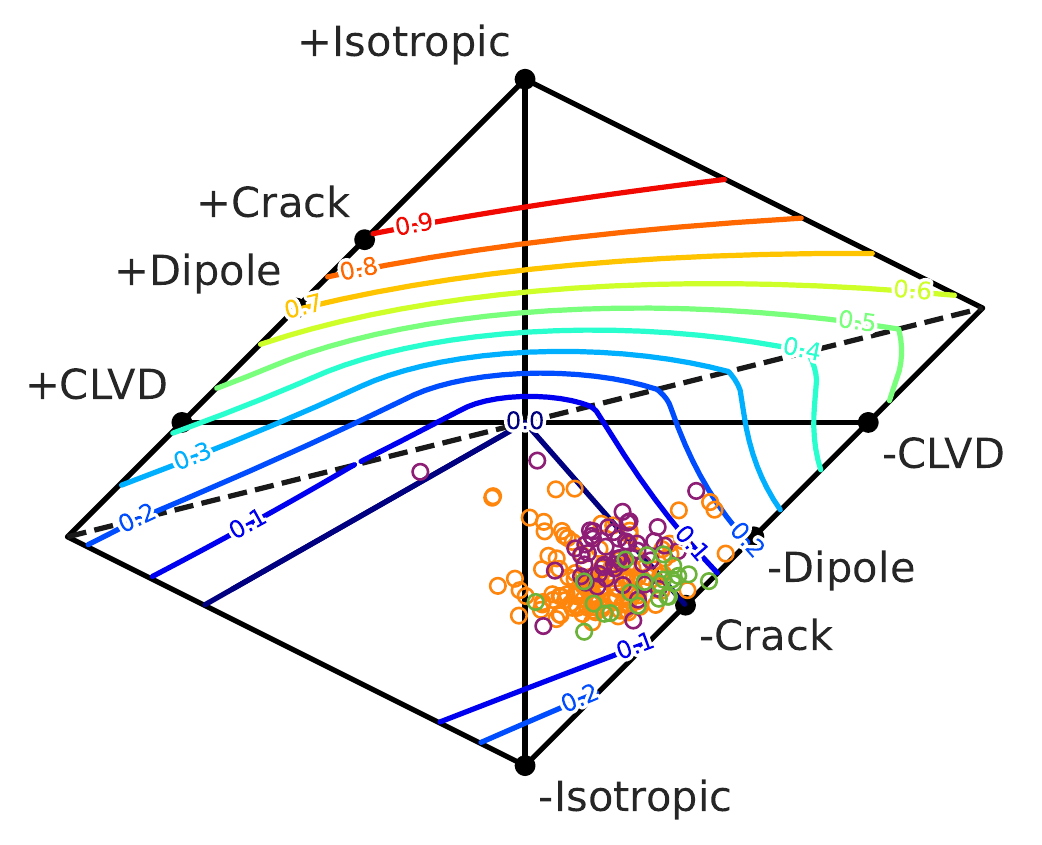}

}
\par\end{flushleft}%
\end{minipage}

\vspace{0.5cm}

\noindent\begin{minipage}[t]{1\textwidth}%
\subfloat[\label{fig:atn-p}]{~~~\includegraphics[bb=0bp 0bp 298bp 239bp,width=0.33\textwidth]{ 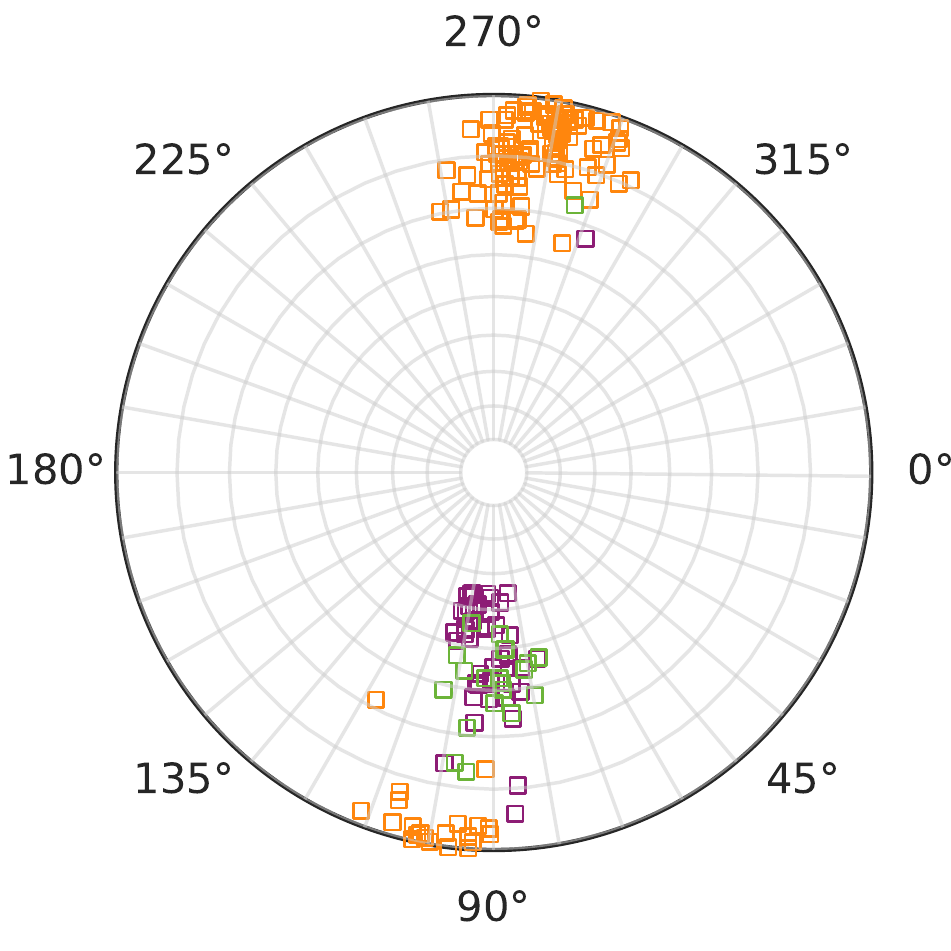}}\subfloat[\label{fig:atn-b}]{~~~\includegraphics[bb=0bp 0bp 298bp 239bp,width=0.33\textwidth]{ 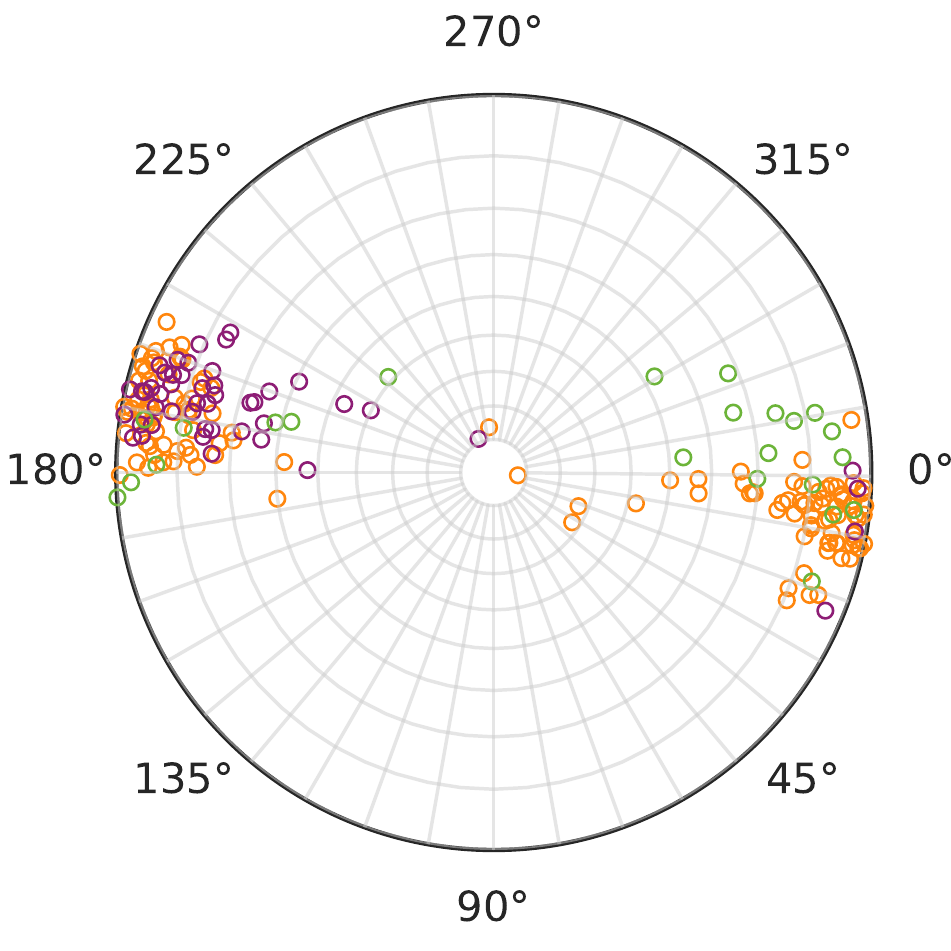}}\subfloat[\label{fig:atn-t}]{~~~\includegraphics[bb=0bp 0bp 298bp 239bp,width=0.33\textwidth]{ 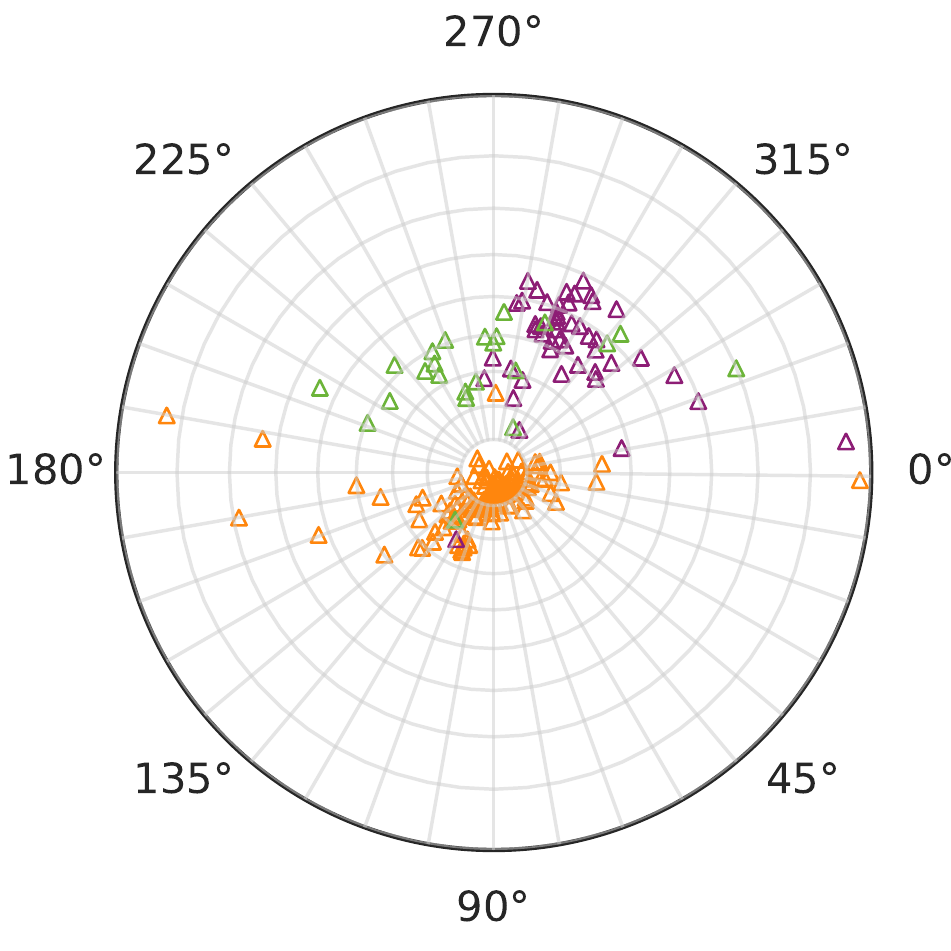}}%
\end{minipage}

\caption{Case study B data summary. (a) View looking west, where the grey wireframes indicate mined-out tunnels and stopes. Seismic events are rendered as beachballs centered at their inferred hypocenter, with red dipoles indicating the $P$-axis direction. Events locating near two tunnels are grouped spatially into clusters A, B, and C. (b)-(c) Plan views of the top and bottom tunnels, respectively, with the direction of \emph{in situ} maximum and intermediate principal compressive stress $\sigma_{1}$ and $\sigma_{2}$, respectively, indicated. (d) View looking south. (e) Hudson source-type plot with colouring according to cluster. The contours indicate amount of non-closing-CDC content as defined in Sec \ref{subsec:Contours}. (f)-(h) Stereonet plots showing the $P$-, $B$-, and $T$-axis orientation, respectively, with colouring according to cluster.\label{fig:(a)-View-looking}}
\end{figure*}

The second catalogue we consider, which is shown in Fig. \ref{fig:(a)-View-looking}, comes from a Western Australian mine. It consists of 204 events from 16 May 2019 to 21 June 2019, which range from $\unit[M_{W}]{-1.4}$ to $\unit[M_{W}]{0.9}$. As can be seen in Figs. \ref{fig:atn-west}-\ref{fig:atn-south}, these events locate to the backs (roofs) of two tunnels that are relatively isolated from stopes or other large excavations (note that these locations are the point of failure initiation). Spatially, they form three clusters: one along the upper tunnel (labeled cluster A) and two along the lower tunnel separated by an intersection (labeled cluster B to the south and cluster C to the north). Figs. \ref{fig:atn-hudson}-\ref{fig:atn-t} show that the moment tensors for all events lie in or near the closing-CDC region (for $\nu=0.23$) and are oriented relatively consistently within each cluster (particularly the $P$-axes).

We have determined a closing-CDC decomposition of each moment tensor following the same three-step procedure outlined in Section \ref{subsec:Case-A}, with the only modification being to the direction of the closing-crack $P$-axis orientation. As noted in Section \ref{subsec:Crush-type-sources}, convergence of the surrounding rockmass into a tunnel is expected to produce a moment tensor with a $P$-axis in the direction of maximum loading orthogonal to the tunnel's axis. Given that they are oriented horizontally at an azimuth of approximately $10^{\circ}$ and both $\sigma_{1}$ and $\sigma_{2}$ in the area are horizontal (orientations given in Figs. \ref{fig:atn-plan-upper} and \ref{fig:atn-plan-lower}), we select decompositions with closing-crack $P$-axis orientations as close as possible to an azimuth of $100^{\circ}$ and plunge of $0^{\circ}$.

\begin{figure}
\subfloat[\label{fig:atn-crush-A}]{\includegraphics[width=0.48\columnwidth]{ 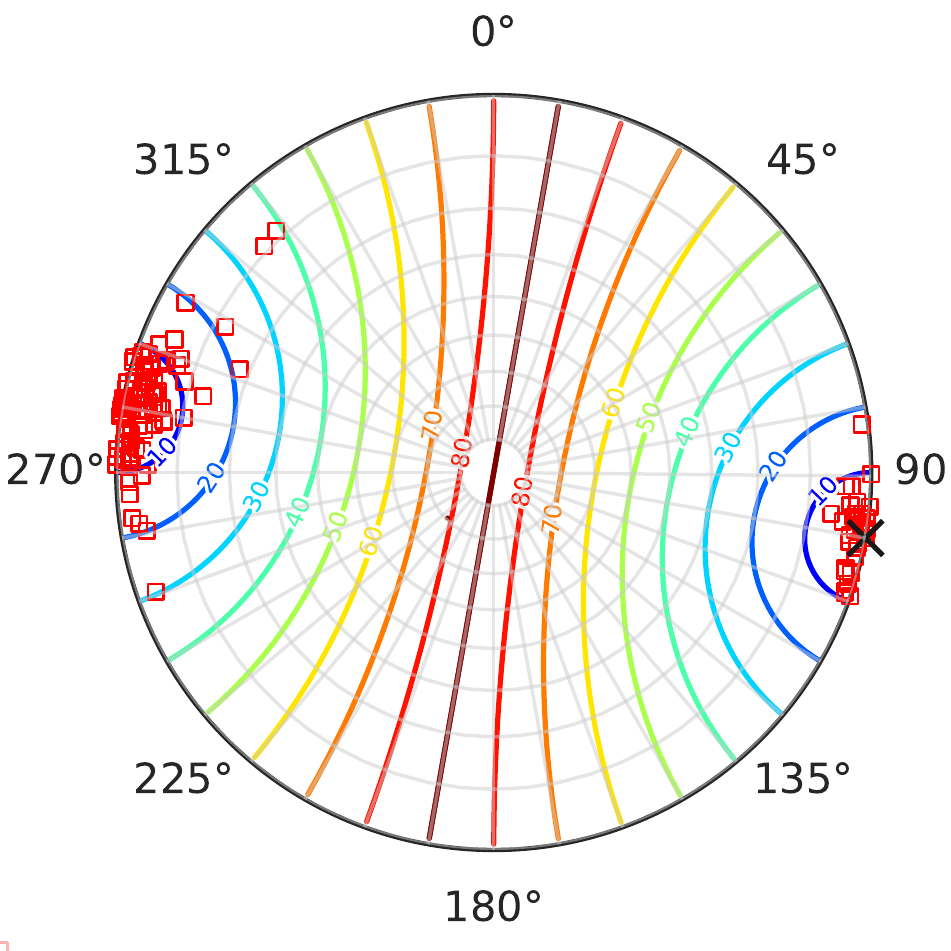}}\subfloat[\label{fig:atn-dc-A}]{\includegraphics[width=0.48\columnwidth]{ 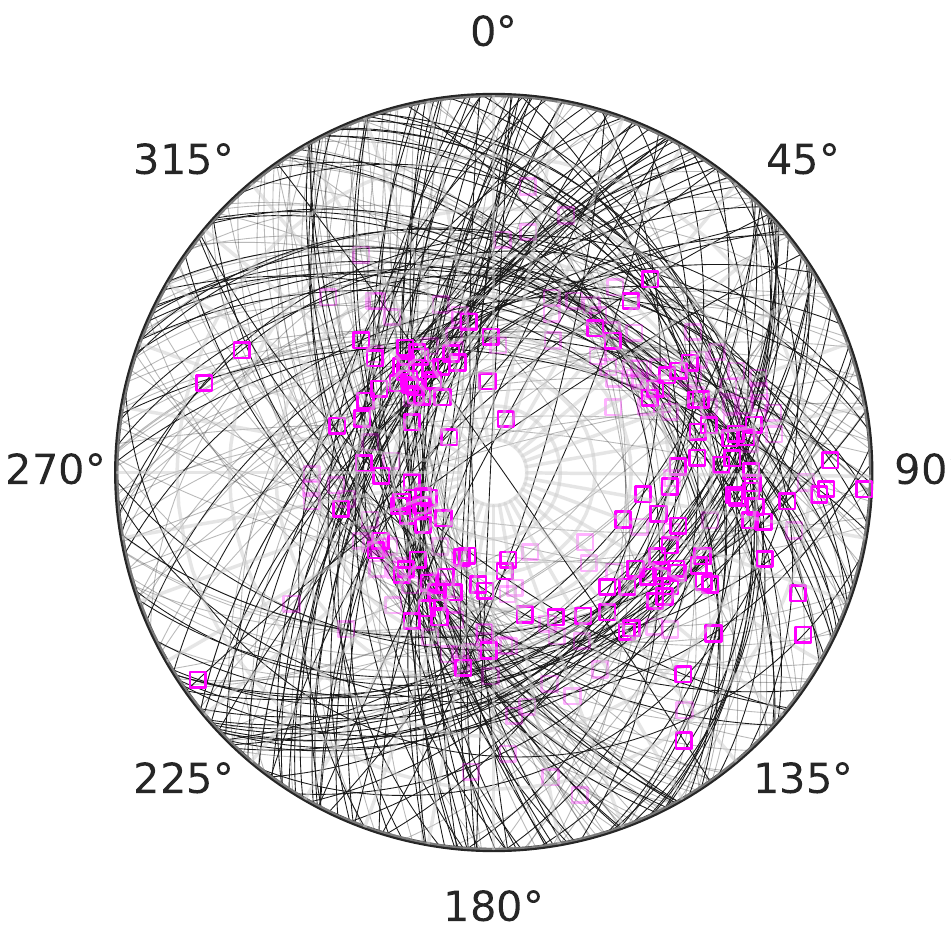}}\vspace{-0.3cm}

\subfloat[\label{fig:atn-crush-B}]{\includegraphics[width=0.48\columnwidth]{ 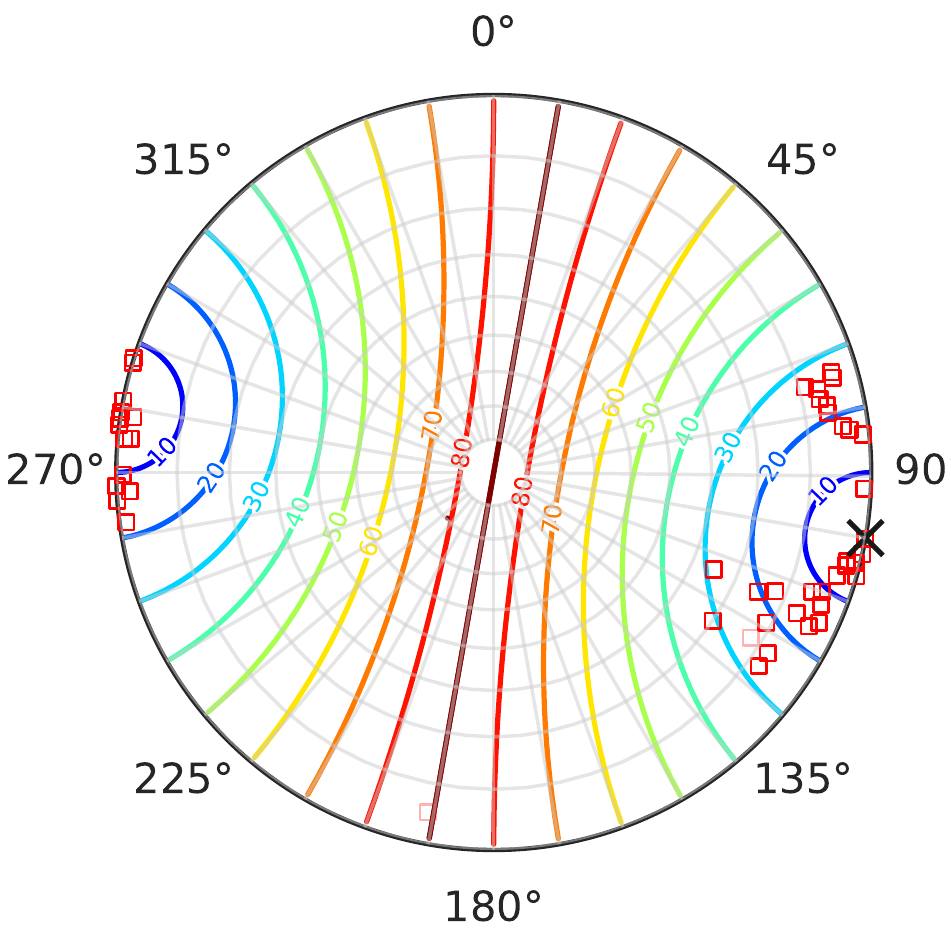}}\subfloat[\label{fig:atn-dc-B}]{\includegraphics[width=0.48\columnwidth]{ 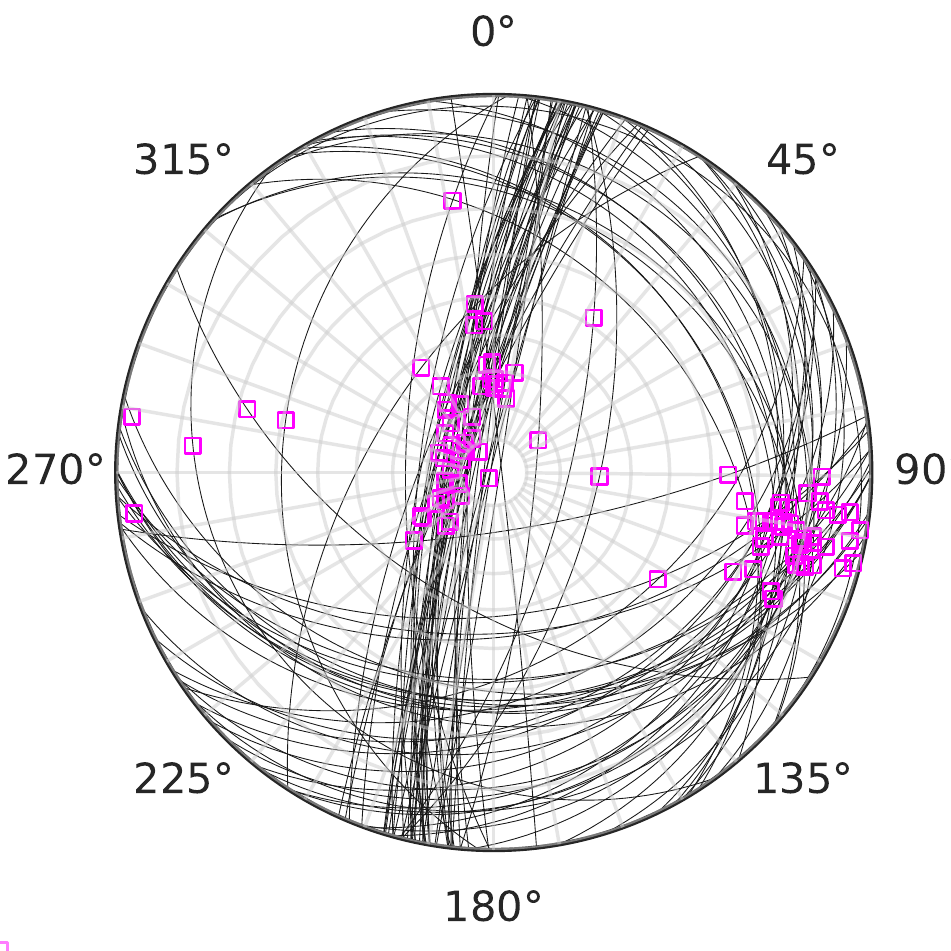}}\vspace{-0.3cm}

\subfloat[\label{fig:atn-crush-C}]{\includegraphics[width=0.48\columnwidth]{ 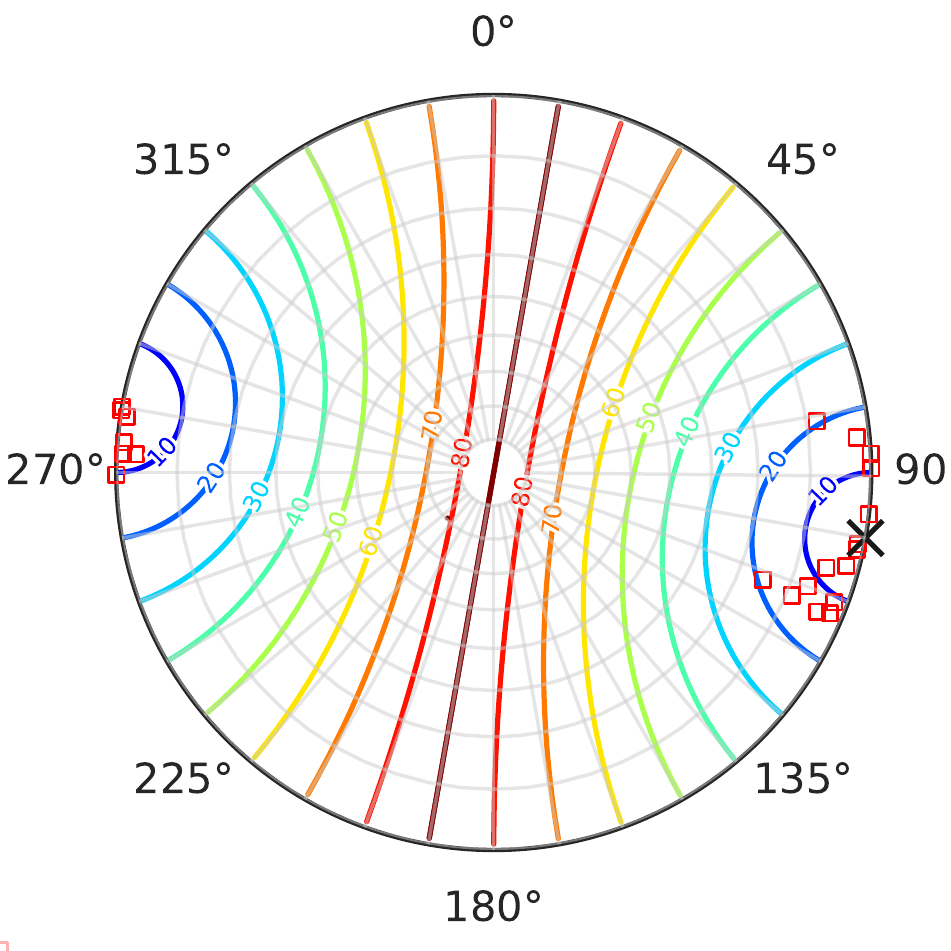}}\subfloat[\label{fig:atn-dc-C}]{\includegraphics[width=0.48\columnwidth]{ 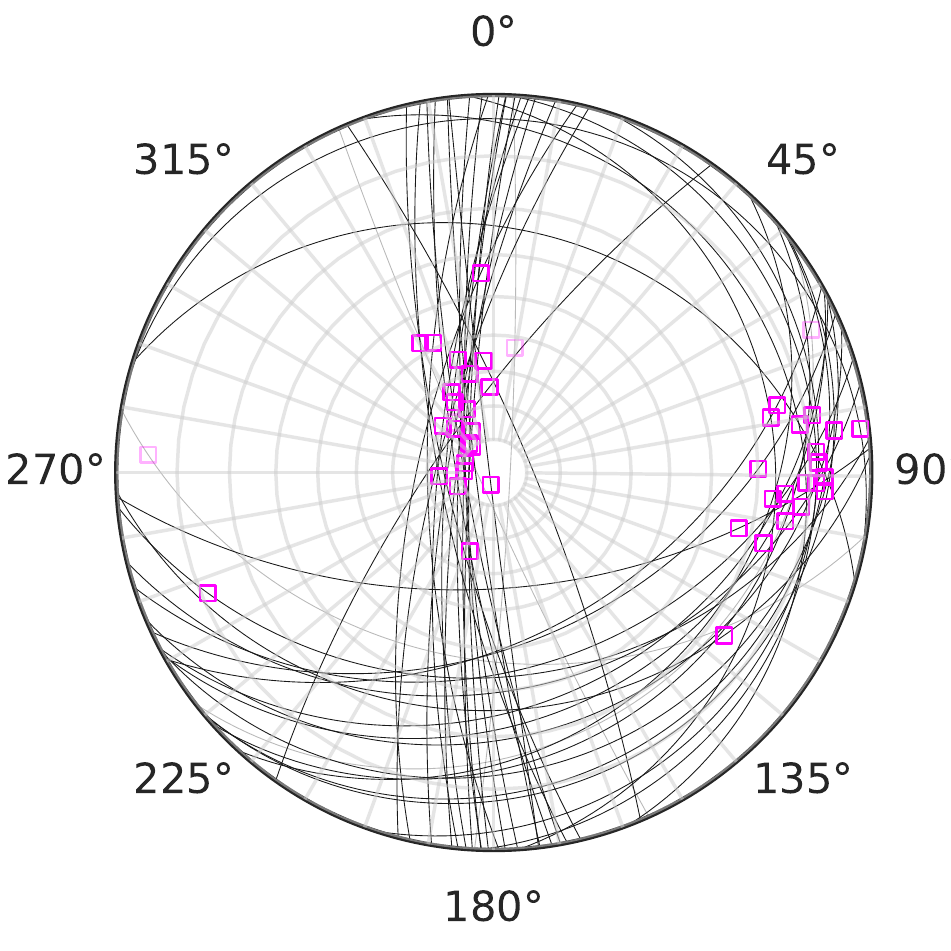}}\vspace{-0.3cm}

\subfloat[\label{fig:atn-scalar-moment-ratios-1}]{\includegraphics[width=0.96\columnwidth]{ 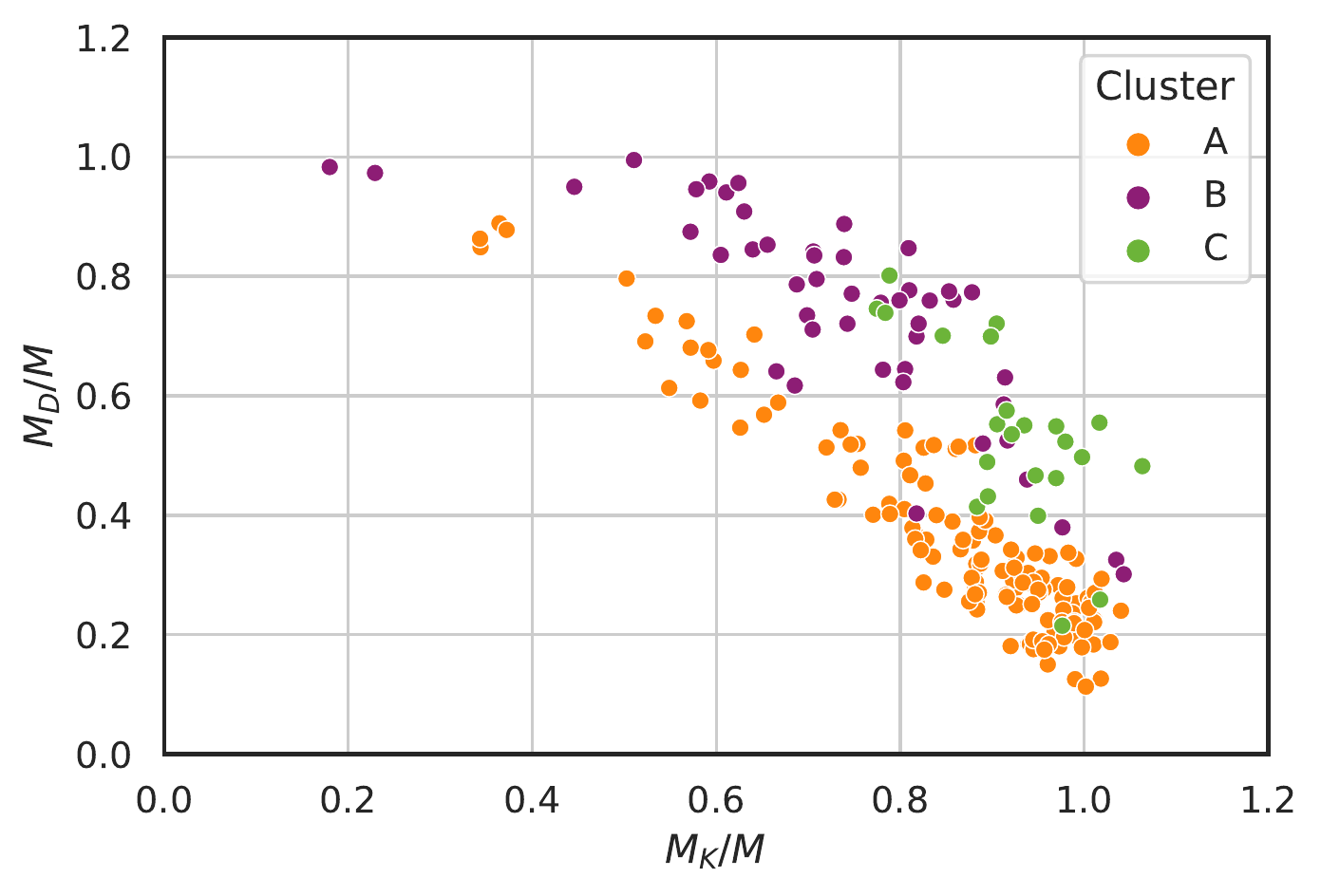}

}

\caption{Case study B decomposition summary. (a) Stereonet of closing-crack $P$-axis orientations for cluster A, which are translucent for $M_{K}/M<0.3$. The contours show the angle to the desired orientation, which is given by the black cross. (b) Stereonet of the DC nodal planes and poles for cluster A, which are translucent for $M_{D}/M<0.3$. (c) Same as (a) for cluster B. (d) Same as (b) for cluster B. (e) Same as (a) for cluster C. (f) Same as (b) for cluster C. (g) Normalised scalar moments coloured by cluster.\label{fig:(a)-Stereonet-of-1}}
\end{figure}

For cluster A, it can be seen in Fig. \ref{fig:atn-crush-A} that for the majority of cases, it was possible to construct a decomposition with a closing-crack $P$-axis close to the desired orientation. Furthermore, as shown in Fig. \ref{fig:atn-scalar-moment-ratios-1}, a large fraction of the decompositions have a relatively small DC component, with approximately half having $M_{D}/M<0.3$. It is reasonable to interpret these sources as episodes of stress fracturing in the back of the tunnel accompanied by convergence of the surrounding rockmass into the excavation. Fig. \ref{fig:atn-dc-A} shows that DC components of those decompositions with $M_{D}/M>0.3$ do not share a common orientation (there are possibly two clusters in these orientations, but it is unclear).

For spatial clusters B and C, it was also possible to determine decompositions with closing-crack $P$-axis close to the desired orientation as shown in Figs. \ref{fig:atn-crush-B} and \ref{fig:atn-crush-C}, respectively. On average, it can be seen from Fig. \ref{fig:atn-scalar-moment-ratios-1}, that these decompositions have more DC content than those found for cluster A (particularly in the case of cluster B). Furthermore, as shown in Figs. \ref{fig:atn-dc-B} and \ref{fig:atn-dc-C}, these DC components largely share a common orientation. It is therefore reasonable to interpret the sources in these clusters as being episodes of subhorizontal or subvertical slipping/shearing in the back of the lower tunnel combined with convergence of the surrounding rockmass.

\section{Conclusion}

\label{sec:Conclusion}

Motivated by the existing classification of mining-induced events into crush- and slip/shear-type sources \citep{ryder1988excess}, we have investigated the closing-CDC decomposition of moment tensors into closing-crack and DC components. In doing this, we have built on the results of \citet{tape2013classical}, who considered the essentially equivalent (mathematically speaking) case of decomposition into opening-crack and DC components. They noted that not every moment tensor permits such a decomposition and expressed the set that do in terms of bounds on the lune. We have translated these bounds in Section \ref{subsec:Hudson} to the source-type plot of \citet{hudson1989source}, which sees more widespread use for historical reasons. 

\citet{tape2013classical} also noted that the moment tensors falling within the closing-CDC region permit an infinite number of decompositions in general. In Section \ref{subsec:Parameterisation}, we have derived an implicit equation that defines this set in terms of the azimuth and plunge of the closing-crack $P$-axis orientation. To determine a finite subset of approximate solutions to this equation, we have made use of the marching-squares algorithm. While this is a relatively robust and performant approach, there is potential for optimisation here (or perhaps even the derivation of parametric solutions). We have also outlined in Section \ref{subsec:Selection} a number of physically motivated criteria that can be used for the selection of a single decomposition from this set.

For moment tensors falling outside the closing-CDC region, we have presented a simple geometric method of determining the closest closing-CDC moment tensor in Section \ref{sec:Approximate-decomposition}, which can then be decomposed. This is a useful tool and allows for the analysis of moment tensors obtained from unconstrained inversion. However, it is not necessarily the case that the closing-CDC moment tensor found by this procedure is the one that gives the best fit to waveforms. Where possible, it may therefore be better to utilise an inversion scheme that is constrained to the closing-CDC region.

As a demonstration of the methods developed, we have analysed two catalogues of mining-induced seismic events in Section \ref{sec:Example-applications}. In both cases, we showed that constraining the expected orientation of the closing-crack's $P$-axis can provide insight into the physical process responsible for the remaining DC component. The information about excavation geometry and stress state near the source that is required to do this is typically readily available, meaning that there is potential for such procedures to be implemented for the routine interpretation of seismic sources at mines.

The physics of the mining-induced sources considered in this paper motivates consideration of only closing-crack content, which can be attributed to excavation closure. However, there are other contexts (such as hydraulic fracturing), where opening-crack content is to be expected. All of the results we have presented for the closing-CDC case can be readily translated to this opening-CDC case. Handling the more general case of an arbitrary (closing or opening) crack is also possible as outlined in Appendix \ref{sec:Arbitrary-crack}. We note that for opening- or arbitrary-CDC decompositions to be useful, relevant selection criteria like those presented in Section \ref{subsec:Selection} would have to be identified for the source processes being considered.

\section*{Acknowledgments}

We are thankful to the mines that provided permission to use their data as presented in Section \ref{sec:Example-applications}.

\section*{Data availability}

The data used in this paper are not publicly available.

\bibliographystyle{/home/ajrigby/Downloads/gji-latex/gji}
\bibliography{cdc}

\begin{thebibliography}{19}
\expandafter\ifx\csname natexlab\endcsname\relax\def\natexlab#1{#1}\fi

\bibitem[Aki \& Richards(2002)]{Aki-Richards-2009}
Aki, K. \& Richards, P., 2002.
\newblock {\it Quantitative seismology\/}, {University Science Books}, 2nd edn.

\bibitem[Ford et~al.(2008)Ford, Dreger, \& Walter]{ford2008source}
Ford, S.~R., Dreger, D.~S., \& Walter, W.~R., 2008.
\newblock Source characterization of the 6 {A}ugust 2007 {C}randall {C}anyon
  {M}ine seismic event in central {U}tah, {\it Seismological Research
  Letters\/}, {\bf 79}(5), 637--644.

\bibitem[Gasperini \& Vannucci(2003)]{gasperini2003fpspack}
Gasperini, P. \& Vannucci, G., 2003.
\newblock {FPSPACK}: a package of {FORTRAN} subroutines to manage earthquake
  focal mechanism data, {\it Computers \& Geosciences\/}, {\bf 29}(7),
  893--901.

\bibitem[Hudson et~al.(1989)Hudson, Pearce, \& Rogers]{hudson1989source}
Hudson, J., Pearce, R., \& Rogers, R., 1989.
\newblock Source type plot for inversion of the moment tensor, {\it Journal of
  Geophysical Research: Solid Earth\/}, {\bf 94}(B1), 765--774.

\bibitem[Julian et~al.(1998)Julian, Miller, \& Foulger]{julian1998non}
Julian, B.~R., Miller, A.~D., \& Foulger, G., 1998.
\newblock Non-double-couple earthquakes 1. {T}heory, {\it Reviews of
  Geophysics\/}, {\bf 36}(4), 525--549.

\bibitem[Knopoff \& Randall(1970)]{knopoff1970compensated}
Knopoff, L. \& Randall, M.~J., 1970.
\newblock The compensated linear-vector dipole: {A} possible mechanism for deep
  earthquakes, {\it Journal of Geophysical Research\/}, {\bf 75}(26),
  4957--4963.

\bibitem[Lorensen \& Cline(1987)]{lorensen1987marching}
Lorensen, W.~E. \& Cline, H.~E., 1987.
\newblock Marching cubes: {A} high resolution 3{D} surface construction
  algorithm, {\it ACM siggraph computer graphics\/}, {\bf 21}(4), 163--169.

\bibitem[Malovichko(2020)]{Malovichko-2020}
Malovichko, D., 2020.
\newblock Description of seismic sources in underground mines: Theory, {\it
  Bulletin of the Seismological Society of America\/}, {\bf 110}.

\bibitem[Malovichko \& Rigby(2022)]{malovichko2022description}
Malovichko, D. \& Rigby, A., 2022.
\newblock Description of seismic sources in underground mines: Dynamic stress
  fracturing around tunnels and strainbursting, {\it arXiv preprint
  arXiv:2205.07379\/}.

\bibitem[Ortlepp(1997)]{ortlepp1997rock}
Ortlepp, W.~D., 1997.
\newblock {\it {R}ock {F}racture and {R}rockbursts: an illustrative study\/},
  vol.~9, South African Institute of Mining and Metallurgy.

\bibitem[Ryder(1988)]{ryder1988excess}
Ryder, J., 1988.
\newblock Excess shear stress in the assessment of geologically hazardous
  situations, {\it Journal of the Southern African Institute of Mining and
  Metallurgy\/}, {\bf 88}(1), 27--39.

\bibitem[Sileny et~al.(2001)Sileny, Psencik, \& Young]{Sileny-2001}
Sileny, J., Psencik, I., \& Young, R.~P., 2001.
\newblock {Point-source inversion neglecting a nearby free surface: simulation
  of the Underground Research Laboratory, Canada}, {\it Geophysical Journal
  International\/}, {\bf 146}, 171--180.

\bibitem[Tape \& Tape(2012{\natexlab{a}})]{tape2012geometric}
Tape, W. \& Tape, C., 2012{\natexlab{a}}.
\newblock A geometric setting for moment tensors, {\it Geophysical Journal
  International\/}, {\bf 190}(1), 476--498.

\bibitem[Tape \& Tape(2012{\natexlab{b}})]{tape2012geometric-comparison}
Tape, W. \& Tape, C., 2012{\natexlab{b}}.
\newblock A geometric comparison of source-type plots for moment tensors, {\it
  Geophysical Journal International\/}, {\bf 190}(1), 499--510.

\bibitem[Tape \& Tape(2013)]{tape2013classical}
Tape, W. \& Tape, C., 2013.
\newblock The classical model for moment tensors, {\it Geophysical Journal
  International\/}, {\bf 195}(3), 1701--1720.

\bibitem[Tape \& Tape(2019)]{tape2019eigenvalue}
Tape, W. \& Tape, C., 2019.
\newblock The eigenvalue lune as a window on moment tensors, {\it Geophysical
  Journal International\/}, {\bf 216}(1), 19--33.

\bibitem[Vavry{\v{c}}uk(2015)]{vavryvcuk2015moment}
Vavry{\v{c}}uk, V., 2015.
\newblock Moment tensor decompositions revisited, {\it Journal of
  Seismology\/}, {\bf 19}(1), 231--252.

\bibitem[Walter et~al.(1997)Walter, Heuze, \& Dodge]{walter1997seismic}
Walter, W.~R., Heuze, F., \& Dodge, D., 1997.
\newblock Seismic {S}ignals from {U}nderground {C}avity {C}ollapses and {O}ther
  {M}ining-{R}elated {F}ailures, Tech. rep., Lawrence Livermore National
  Lab.(LLNL), Livermore, CA (United States).

\bibitem[Wong \& McGarr(1991)]{wong1991implosional}
Wong, I. \& McGarr, A., 1991.
\newblock Implosional failure in mining-induced seismicity: a critical review,
  in {\em International Journal of Rock Mechanics and Mining Sciences and
  Geomechanics Abstracts\/}, vol.~28, pp. A397--A397, Elsevier Science.

\end{thebibliography}

\appendix

\section{Arbitrary crack}

\label{sec:Arbitrary-crack}

Sources with opening-crack content are less relevant in the context of mining-induced seismicity than than those with closing-crack content as considered in the main text. However, they are potentially of interest in different contexts, so we give here an overview of how the closing-CDC decomposition can be extended to an arbitrary crack (opening or closing). For the sake of brevity, we present results only on the lune, but they can be translated to the Hudson plot in a fairly straightforward manner.

\subsection{Source model}

An arbitrary-CDC source has a moment tensor of the form has a moment tensor of the form
\begin{equation}
\mathbf{M}=\mathbf{M}^{K}+\mathbf{M}^{D},\label{eq:source-model-1-1}
\end{equation}
where $\mathbf{M}^{K}$ is either a closing-crack moment tensor $\mathbf{M}^{K^{-}}$ or opening crack moment tensor $\mathbf{M}^{K^{+}}=-\mathbf{M}^{K^{-}}$. The opening-crack can be parameterised as $\mathbf{M}^{K^{+}}(M^{K},\phi,\beta)$ in terms of its scalar moment $M^{K}$ and the azimuth $\phi$ and plunge $\beta$ of the $T$-axis {[}similar to eq. \eqref{eq:closing-crack-components}{]}.

\subsection{Source-type bounds}

\begin{figure}
\begin{centering}
\includegraphics[bb=30bp 0bp 194bp 235bp,clip,scale=0.75]{ 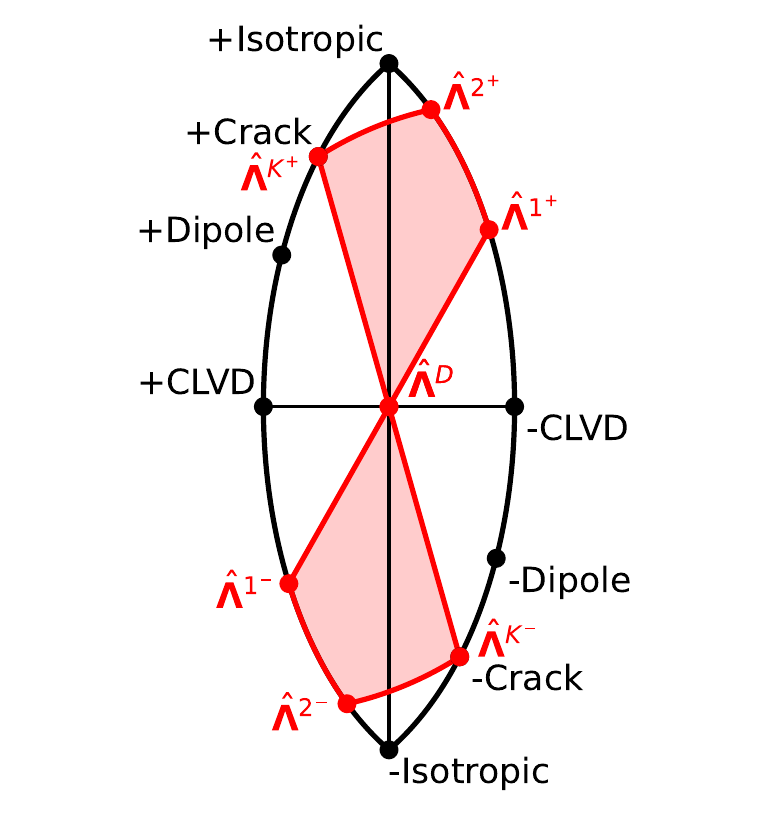}
\par\end{centering}
\caption{\label{fig:lune-cdc-region-both}Region on the fundamental lune for which corresponding moment tensors permit an arbitrary-CDC decomposition with $\nu=0.25$.}
\end{figure}

The region $\mathbb{L}_{\mathrm{CDC}}$ of the lune permitting an arbitrary-CDC decomposition is shown in Fig. \ref{fig:lune-cdc-region-both}. It is the union of $\mathbb{L}_{\mathrm{CDC}^{-}}$ as defined in eq. \eqref{eq:cdc-region-def} and the region $\mathbb{L}_{\mathrm{CDC}^{+}}$ that permits decomposition into opening-crack and DC components. By defining 
\begin{equation}
r_{D}(\lambda_{1},\lambda_{2},\lambda_{3})=(-\lambda_{3},-\lambda_{2},-\lambda_{1}),
\end{equation}
which is a rotation by $180^{\circ}$ around $\hat{\boldsymbol{\Lambda}}^{D}$, the vertices of $\mathbb{L}_{\mathrm{CDC}^{+}}=r_{D}(\mathbb{L}_{\mathrm{CDC}^{-}})$ can be written as $\hat{\boldsymbol{\Lambda}}^{D}$, $\hat{\boldsymbol{\Lambda}}^{K^{+}}=r_{D}(\hat{\boldsymbol{\Lambda}}^{K^{-}})$, $\hat{\boldsymbol{\Lambda}}^{1^{+}}=r_{D}(\hat{\boldsymbol{\Lambda}}^{1^{-}})$, and $\hat{\boldsymbol{\Lambda}}^{2^{+}}=r_{D}(\hat{\boldsymbol{\Lambda}}^{2^{-}})$. A more useful definition for determining whether a given $\hat{\boldsymbol{\Lambda}}$ belongs to $\mathbb{L}_{\mathrm{CDC}^{+}}$ is
\begin{equation}
\mathbb{L}_{\mathrm{CDC}^{+}}=\{\hat{\boldsymbol{\Lambda}}\in\text{\ensuremath{\mathbb{L}}}:r_{D}(\hat{\boldsymbol{\Lambda}})\in\mathbb{L}_{\mathrm{CDC}^{-}}\}.
\end{equation}

\subsection{Decomposition generation/selection}

If $\hat{\boldsymbol{\Lambda}}\in\mathbb{L}_{\mathrm{CDC}^{-}}$, then the set of possible decompositions is defined according to eq. \eqref{eq:det-constraint}. If instead $\hat{\boldsymbol{\Lambda}}\in\mathbb{L}_{\mathrm{CDC}^{+}}$, then the opening-crack $T$-axis orientation will satisfy
\begin{equation}
\det\left[\mathbf{M}-\mathbf{M}^{K^{+}}\left(\frac{\mathrm{tr(}\mathbf{M})}{(\nu+1)\alpha_{v}},\phi,\beta\right)\right]=0.\label{eq:det-constraint-1}
\end{equation}
The methods used for selecting a single decomposition from this set will depend on the nature of the source process being considered.

\subsection{Nearest CDC moment tensor}

\begin{figure}
\begin{centering}
\includegraphics[bb=30bp 0bp 194bp 235bp,clip,scale=0.75]{ 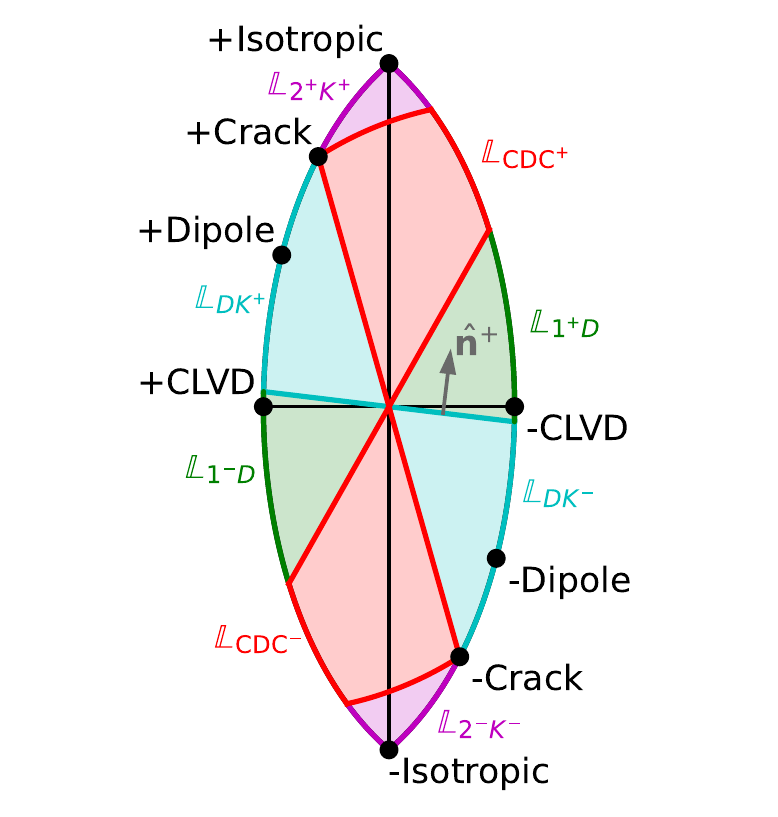}
\par\end{centering}
\caption{\label{fig:lune-regions-both}Partition of the lune into the regions relevant to an arbitrary-CDC decomposition for $\nu=0.25$. }
\end{figure}

If $\hat{\boldsymbol{\Lambda}}\notin\mathbb{L}_{\mathrm{CDC}}$, then the closest $\hat{\boldsymbol{\Lambda}}^{\mathrm{CDC}}\in\mathbb{L}_{\mathrm{CDC}}$ depends on which of the regions shown in Fig. \ref{fig:lune-regions-both} $\hat{\boldsymbol{\Lambda}}$ belongs to. $\mathbb{L}_{2^{-}K^{-}}$ is the same as defined in eq. \eqref{eq:lune-partition}. The regions 
\begin{align}
\mathbb{L}_{DK^{-}} & =\{\hat{\boldsymbol{\Lambda}}\in\mathbb{L}:\hat{\mathbf{n}}^{+}\cdot\hat{\boldsymbol{\Lambda}}<0,\hat{\mathbf{n}}^{DK}\cdot\hat{\boldsymbol{\Lambda}}>0\},\nonumber \\
\mathbb{L}_{1^{-}D} & =\{\hat{\boldsymbol{\Lambda}}\in\mathbb{L}:\hat{\mathbf{n}}^{+}\cdot\hat{\boldsymbol{\Lambda}}<0,\hat{\mathbf{n}}^{1D}\cdot\hat{\boldsymbol{\Lambda}}>0\},
\end{align}
are modified slightly from their definitions in eq. \eqref{eq:lune-partition}, now depending on 
\begin{equation}
\hat{\mathbf{n}}^{+}=p_{\mathbb{S}}(\hat{\mathbf{n}}^{DK}+\hat{\mathbf{n}}^{1D}).
\end{equation}
The remaining regions are their rotations: $\mathbb{L}_{2^{+}K^{+}}=r_{D}(\mathbb{L}_{2^{-}K^{-}})$, $\mathbb{L}_{DK^{+}}=r_{D}(\mathbb{L}_{DK^{-}})$, and $\mathbb{L}_{1^{+}D}=r_{D}(\mathbb{L}_{1^{-}D})$. In terms of these regions, 
\begin{equation}
\hat{\boldsymbol{\Lambda}}^{\mathrm{CDC}}=\begin{cases}
\hat{\boldsymbol{\Lambda}} & \mathrm{if\,}\hat{\boldsymbol{\Lambda}}\in\mathbb{L}_{\mathrm{CDC}},\\
p_{\mathbb{S}}(\hat{\boldsymbol{\Lambda}}-[\hat{\boldsymbol{\Lambda}}\cdot\hat{\mathbf{n}}^{DK}]\hat{\mathbf{n}}^{DK}) & \mathrm{if\,}\hat{\boldsymbol{\Lambda}}\in\mathbb{L}_{D^{\pm}K^{\pm}},\\
p_{\mathbb{S}}(\hat{\boldsymbol{\Lambda}}-[\hat{\boldsymbol{\Lambda}}\cdot\hat{\mathbf{n}}^{1D}]\hat{\mathbf{n}}^{1D}) & \mathrm{if\,}\hat{\boldsymbol{\Lambda}}\in\mathbb{L}_{1^{\pm}D^{\pm}},\\
p_{\mathbb{S}}(\hat{\boldsymbol{\Lambda}}-[\hat{\boldsymbol{\Lambda}}\cdot\hat{\mathbf{n}}^{2^{\pm}K^{\pm}}]\hat{\mathbf{n}}^{2^{\pm}K^{\pm}}) & \mathrm{if\,}\hat{\boldsymbol{\Lambda}}\in\mathbb{L}_{2^{\pm}K^{\pm}},
\end{cases}
\end{equation}
where $\hat{\mathbf{n}}^{2^{+}K^{+}}=r_{D}(\hat{\mathbf{n}}^{2^{-}K^{-}})$.

\subsection{Non-CDC content contours}

\begin{figure}
\begin{centering}
\includegraphics[bb=30bp 0bp 194bp 235bp,clip,scale=0.75]{ 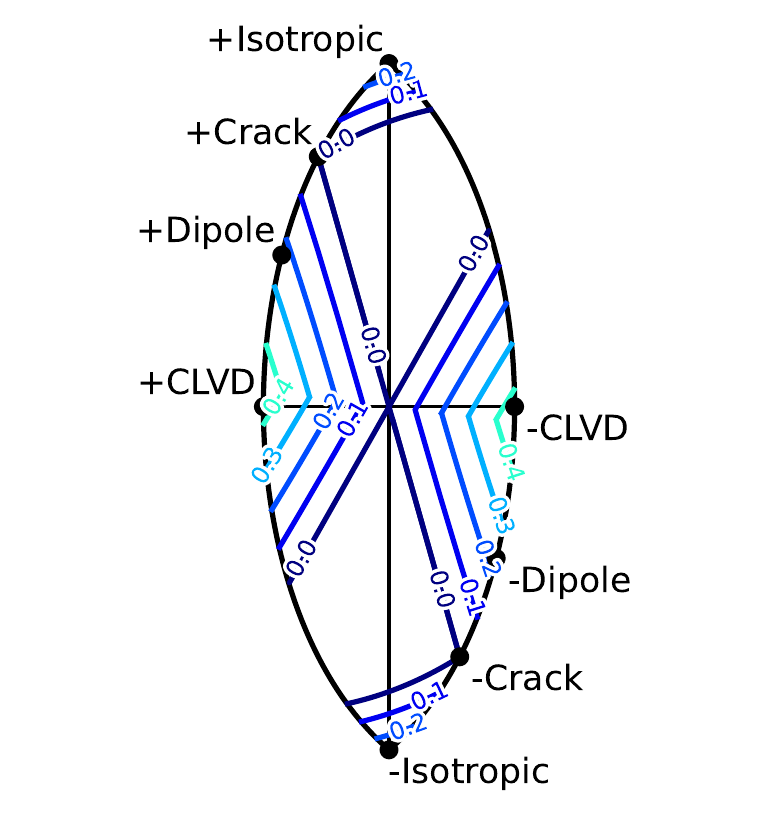}
\par\end{centering}
\caption{\label{fig:lune-contours-both}Isolines of constant non-CDC amount on the lune for $\nu=0.25$.}
\end{figure}

Isolines of constant non-CDC content $\gamma_{\mathrm{CDC}}$ {[}defined similarly to eq. \eqref{eq:non-cdc-amount}{]} are shown in Fig. \ref{fig:lune-contours-both}. Geometrically, these are again composed of small-circle segments, which can be written in terms of plane intersections as
\begin{equation}
\begin{cases}
\hat{\boldsymbol{\Lambda}}\cdot\hat{\mathbf{n}}^{DK}\pm\gamma_{\mathrm{CDC}}=0 & \mathrm{if}\,\hat{\boldsymbol{\Lambda}}\in\mathbb{L}_{DK^{\pm}},\\
\hat{\boldsymbol{\Lambda}}\cdot\hat{\mathbf{n}}^{1D}\pm\gamma_{\mathrm{CDC}}=0 & \mathrm{if}\,\hat{\boldsymbol{\Lambda}}\in\mathbb{L}_{1^{\pm}D},\\
\hat{\boldsymbol{\Lambda}}\cdot\hat{\mathbf{n}}^{2^{\pm}K^{\pm}}\mp\gamma_{\mathrm{CDC}}=0 & \mathrm{if}\,\hat{\boldsymbol{\Lambda}}\in\mathbb{L}_{2^{\pm}K^{\pm}}.
\end{cases}
\end{equation}

\end{document}